\documentclass[usenatbib,useAMS]{mn2e}
\usepackage{graphicx}
\usepackage{amssymb,amsmath,mathrsfs}

\newcommand{\aap}{A\&A}%
\newcommand{\aj}{AJ}%
\newcommand{\apj}{ApJ}%
\newcommand{\apjl}{ApJ}%
\newcommand{\apjs}{ApJS}%
\newcommand{\araa}{ARA\&A}
\newcommand{\mnras}{MNRAS}%
\newcommand\Rein{R_{\rm Ein}}
\newcommand\Mein{M_{\rm Ein}}
\newcommand{\vect}[1]{\boldsymbol{#1}}

\begin{document}

\title{Substructure in the lens HE 0435$-$1223}

\author[Fadely \& Keeton]{Ross Fadely$^{1,2}$ \& Charles R. Keeton$^{2}$ \\
$^1$(Current) Department of Astronomy, Haverford College, 370 Lancaster Ave., Haverford, PA 19041 USA; rfadely@haverford.edu \\
$^2$Department of Physics and Astronomy, Rutgers, the State University of New Jersey, 136 Frelinghuysen Road, Piscataway, NJ \\ 08854 USA; keeton@physics.rutgers.edu
}

\maketitle

%=====================================================================
\begin{abstract}

We investigate the properties of dark matter substructure in the gravitational lens HE 0435$-$1223 ($z_l=0.455$) via its effects on the positions and flux ratios of the quadruply-imaged background quasar ($z_s=1.689$).  We start with a smooth mass model, add individual, truncated isothermal clumps near the lensed images, and use the Bayesian evidence to compare the quality of different models.  Compared with smooth models, models with at least one clump near image A are strongly favoured.  The mass of this clump within its Einstein radius is $\log_{10}(\Mein^A)=7.65^{+0.87}_{-0.84}$ (in units of $h_{70}^{-1}\,M_\odot$).  The Bayesian evidence provides weaker support for a second clump near image B, with $\log_{10}(\Mein^B)=6.55^{+1.01}_{-1.51}$.  We also examine models with a full population of substructure, and find the mass fraction in substructure at the Einstein radius to be $f_{\rm sub} \ga 0.00077$, assuming the total clump masses follow a mass function $dN/dM \propto M^{-1.9}$ over the range $M=10^7$--$10^{10}\,M_\odot$.  Few-clump and population models produce similar Bayesian evidence values, so neither type of model is objectively favoured.

\end{abstract}

%=====================================================================
\section{Introduction}
\label{sec:intro}
%=====================================================================

A tension has arisen between the cold dark matter (CDM) paradigm and certain astronomical observations.  On the theory side, N-body simulations have reached consensus that galaxy-scale dark matter halos should contain many bound subhalos that follow a power-law mass function, $dN/dM \propto M^\alpha$.  Probing from $\sim\!10^{10}\,M_\odot$ down to $\sim\!10^4\,M_\odot$, simulations such as \textit{Via Lactea} \citep{diemand07a} and \textit{Aquarius} \citep{springel08a} predict a mass function slope of $\alpha \approx -1.9$ and a fractional amount of substructure in the vicinity of 8--11\% (depending to some extent on resolution).

Observationally, however, the prediction of dark matter substructure has not been confirmed.  Various surveys have sought to characterise the abundance, masses, and spatial distribution of low-mass galaxies in the Local Group \citep[e.g.,][]{simonj07a,kalirai10a}.  Before 2005, only the 11 most massive and luminous Milky Way satellites had been found \citep{mateo98a}.  After 2005, the Sloan Digital Sky Survey \citep{york00a} made it possible to detect extremely faint satellites \citep[e.g.,][]{willman05a,irwin07a,liuc08a,belokurov09a,belokurov10a}.  These ``ultra-faint'' dwarfs, with absolute magnitudes as low as $M_V\sim -2$, have more than doubled the number of Milky Way satellites to $\sim25$ \citep[for a current list, see][]{wadepuhl10a}.  Yet, despite this dramatic leap forward, the number of satellites still falls severely short of the hundreds predicted by N-body simulations \citep{klypin99a,moore99a}.

A clear contributor to this disparity is the lack of a complete and thorough survey of the local volume.  Indeed, while a huge improvement over previous surveys, the SDSS is limited in both sky coverage ($\sim$1/5 of the sky) and depth ($g<22.2$).  Attempts to account for these limitations suggest that a volumetrically complete survey will find many more satellites and may eliminate the problem altogether \citep{tollerud08a}.  Those estimates depend, however, on extrapolations from the currently known population, and it is quite plausible that even a complete survey will not find all the predicted satellites.  If so, any remaining discrepancy between theoretical predictions and observations will presumably be attributed to the intrinsic luminosities of low-mass dwarfs.  Satellites with total mass $\la 10^7 M_\odot$ can experience suppressed or even quenched star formation \citep[e.g.,][]{strigari07a,maccio10a}.  Cosmic reionisation, UV photo-evaporation, ram pressure or tidal stripping, supernovae, and cosmic rays may all play a role in hampering the conditions for star formation \citep{gnedin00a,scannapieco01a,strigari07a,madau08a,mashchenko08a,maccio10a,penarrubia10a,wadepuhl10a}.  While the precise mechanisms are still debated, the plausibility of such arguments points to a large population of ``dark dwarfs'', whose luminosities are so low that they will elude traditional observational techniques.

Intriguingly, while local observations of satellite galaxies seem to fall short of CDM predictions, measurements in more distant galaxies exhibit the opposite conflict.  Sensitive to mass alone, strong gravitational lensing provides a unique tool to detect low-mass subhalos in cosmologically distant galaxies, regardless of their luminosities \citep[e.g.,][]{dalal02a,vegetti10b}.  On large angular scales ($\sim\!1''$), the bulk properties of multiply-imaged quasars are determined by the macroscopic mass distribution of the lens galaxy and its surrounding environment.  Upon detailed 
inspection, however, the properties of the images may be perturbed by small-scale structure in the mass distribution \citep{maos98a,chibam02a,metcalf01a,dalal02a,metcalf02a,bradac02a, koopmans02a,chenj07a,keeton09a,keeton09b}.  Thus, with the positions, flux ratios, and time delays of lensed images we may be able to measure the properties of small-scale structure in lens galaxies.

Currently, some of the best constraints on dark matter substructure (outside of the Local Group) come from the analysis of ``anomalous'' flux ratios in four-image gravitational lenses.  Many lenses have flux ratios that violate universal relations predicted for smooth mass models \citep{keeton03a,keeton05a}.  Performing a statistical analysis of seven lenses, \citet{dalal02a} found the mass fraction in substructure to be $0.006<f_{\rm sub}<0.07$ (90\% confidence) at the Einstein radii of the lenses.  This stands in contrast to CDM predictions, which yield $f_{\rm sub} \sim 0.001$--0.003 at similar projected radii \citep{maos04a,amara06a,maccio06b,maccio06a}.  In particular, \citet{xu10a} recently found that N-body simulations predict $f_{\rm sub} \sim 0.002$ at typical Einstein radii even when considering other sources of small-scale structure beyond dark matter substructure (e.g., globular clusters, stellar streams).

Observational constraints, therefore, seem at odds.  Tallies of Milky Way satellites seem to indicate a dearth of substructure, while lensing points to a surplus.  Confronting this on the lensing side, there is great interest in expanding both the list of observables and the sample of lenses used to probe substructure.  For example, infrared observations of lenses have begun to increase the number of quasar lenses available for flux ratio studies \citep[e.g.,][]{chibam05a,macleodc09a,minezaki09a,fadely11a}.  Image positions \citep{chenj07a} and time delays \citep{keeton09a} can complement flux ratios by providing different sensitivity to substructure in quasar lenses.  Also, Einstein ring images offer a new way to probe substructure in galaxy-galaxy strong lenses \citep{vegetti09a,vegetti10a, vegetti10b}.  In particular, \citet{vegetti10b} recently used a Bayesian analysis to infer $f_{\rm sub} = 0.0215^{+0.0201}_{-0.0125}$ in the lens SDSS J0946+1006, assuming $\alpha=-1.9\pm0.1$ for a mass range from $M_{\rm total}=10^{6.6}\,M_\odot$ to $10^{9.6}\,M_\odot$.

\begin{figure}
\centering
\includegraphics[clip=true, trim=0cm 0cm 0.cm 0.cm,width=8cm]{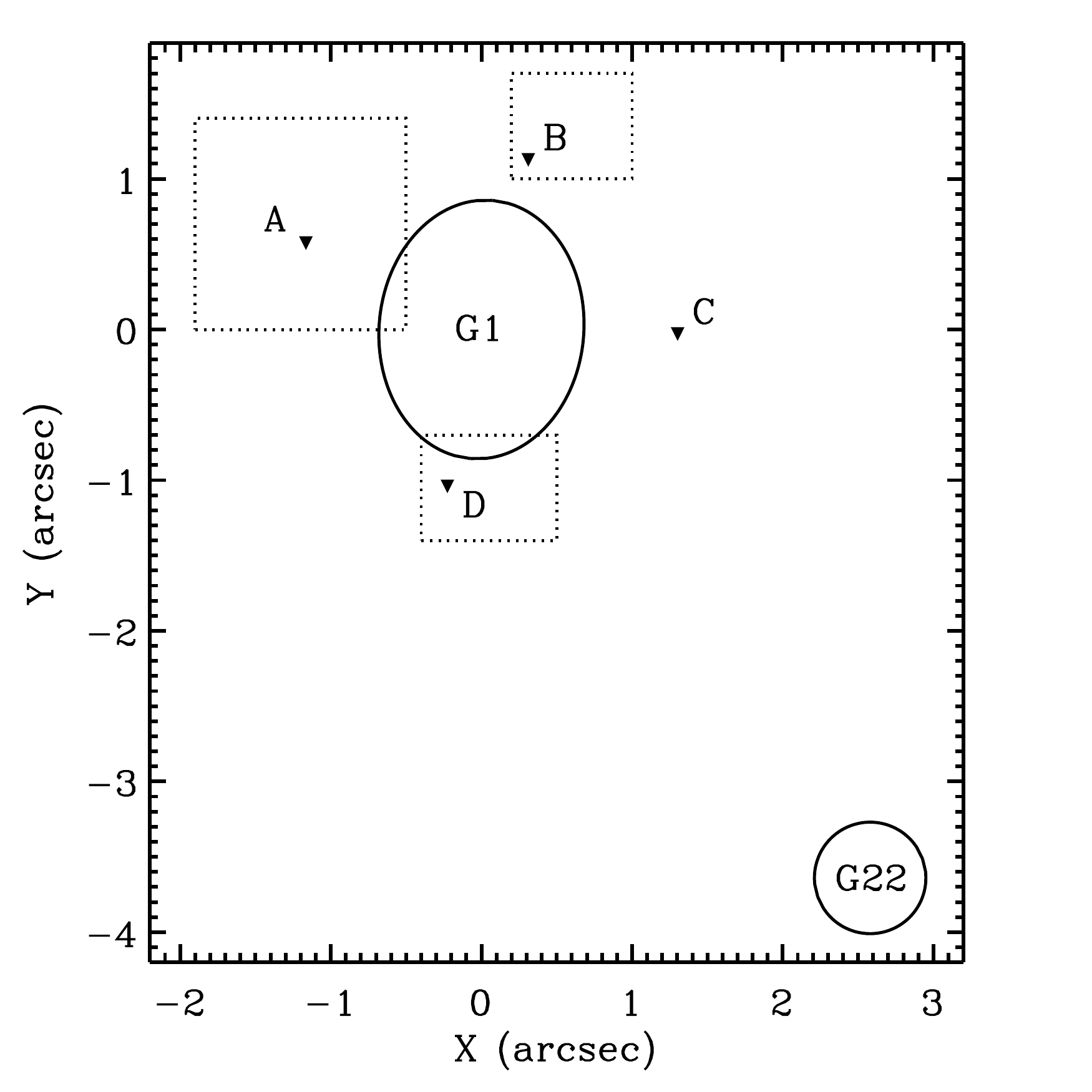}
\caption{
Schematic diagram of the lens HE 0435-1223.  The size of ellipses depicting the main lens G1 and nearby galaxy G22 are set to the effective radii measured in HST images.  For G22, no measurement of the ellipticity is given in the literature.  Dotted boxes present the spatial regions used as priors for the clump positions in Section \ref{sec:few}, after an initial MCMC exploration.
}\label{fig:diagram}
\end{figure}

In this paper we investigate the properties of the four image gravitational lens HE 0435$-$1223 
(hereafter HE0435), selected for its relatively bright ($F160W<18.1$) and well separated ($2.4''$) images (see Fig.~\ref{fig:diagram}).  Since its discovery \citep{wisotzki02a}, HE0435 has been extensively studied using ground- and space-based observations.  From the ground, optical spectroscopy provided early evidence for stellar microlensing and against significant differential dust extinction in the lens \citep{wisotzki03a}.  More recently, optical monitoring has quantified the intrinsic and microlensing variability, and also revealed the time delays between images \citep{kochanek06a,courbin10a}.  \textit{Hubble Space Telescope} imaging provided photometric evidence that the lens lies in an overdense environment \citep{morgann05a}, and pencil-beam redshift surveys have confirmed the presence of a group of galaxies surrounding the lens \citep{wongkc11a}.  A galaxy lying near the lens on the sky \citep[labeled G22 by][]{morgann05a} seems to be important for reproducing the lensed images \citep{kochanek06a}.  Using the available data, including new near-infrared photometry \citep{fadely11a}, we examine the mass distribution of HE0435 and pay particular attention to any evidence for substructure.  New in our analysis is the use of both individual-clump and population-based simulations of substructure, which allow us both to constrain the masses of clumps near the images and to connect them to the broader substructure population.  Not included in our analysis are effects from small mass halos along the line of sight.  While such structures may produce millelensing effects similar to subhalos within lens galaxies, their ultimate importance is still debated \citep[e.g.,][]{chen03,metcalf05b}.  Where necessary we assume a flat cosmology with $\Omega_m=0.27$ and $H_0=70.4\,{\rm km\, s^{-1}\, Mpc^{-1}}$, which is similar to the mean WMAP+BAO+$H_0$ values presented in \citet{komatsu11a}.

%=====================================================================
\section{Constraints}
\label{sec:constraints}
%=====================================================================
 
\begin{table*}
\begin{tabular}{cccccc}
\hline
\multicolumn{6}{c}{Images} \\
\hline
 & \multicolumn{2}{c}{Position ($''$)} & $R$-band flux & $K$-band flux & $L'$-band flux\\
\hline 
Image A & $-1.165\pm0.003$ & $\phantom{-}0.573\pm0.003$ & $1.751\pm0.098$ & $1.837\pm0.086$ & $1.706\pm0.085$  \\
Image B & $\phantom{-}0.311\pm0.004$  & $\phantom{-}1.126\pm0.004$  & $0.998\pm0.037$ & $1.271\pm0.063$ & $0.991\pm0.065$ \\
Image C & $\phantom{-}1.302\pm0.005$  & $-0.030\pm0.005$ & $\equiv 1.0$ & $\equiv 1.0$ & $\equiv 1.0$ \\
Image D & $-0.226\pm0.003$ & $-1.041\pm0.003$ & $0.851\pm0.049$ & $0.745\pm0.049$ & $0.809\pm0.090$ \\
\hline
\multicolumn{6}{c}{Galaxies} \\
\hline
 & \multicolumn{2}{c}{Position ($''$)} & F555W (mag) & F814W (mag) & F160W (mag) \\
\hline
 G1 & $\phantom{'}\equiv 0.0\pm0.002$ & $\phantom{'}\equiv 0.0\pm0.002$ & $21.55\pm0.13$ & $18.85\pm0.13$ & $16.86\pm0.04$ \\ 
 G22 & $\phantom{,'}2.585\pm0.005$ & $-3.637\pm0.005\,$ & $22.25\pm0.04$ & $21.26\pm0.01$ & $\sim$18.8 \\
 \hline
\end{tabular}
\caption{
HE0435 constraints.  The positions and $R$-band photometry of the images are from \citet{kochanek06a}.  The $R$-band flux ratios reflect the mean and standard deviation from light curve monitoring, and include scatter from intrinsic and microlensing variability.  The $K$ and $L'$-band photometry of the images are from \citet{fadely11a}.  The data for the lens galaxy G1, and the F160W magnitude of the neighbor galaxy G22, are from \citet{kochanek06a}, while the remaining data for G22 are from \citet{morgann05a}.
}\label{tab:data}
\end{table*}

Out of the previous observations of HE0435 we must select the data we seek to fit.  The chosen data should provide valuable constraints on the lens mass distribution, be practical to use, and permit a straightforward interpretation.  The optimal astrometric data are the HST-derived centroids of the lensed images and the main lens galaxy, G1 \citep{kochanek06a}, along with the position of the neighboring galaxy, G22 \citep{morgann05a}.  The redshift of G22 is not known \citep[see][]{wongkc11a}; we assume this galaxy lies at the same redshift as G1.  The data are summarised in Table \ref{tab:data}.

\begin{figure}
\centering
\includegraphics[clip=true, trim=0cm 0cm 0.cm 0.cm,width=8cm]{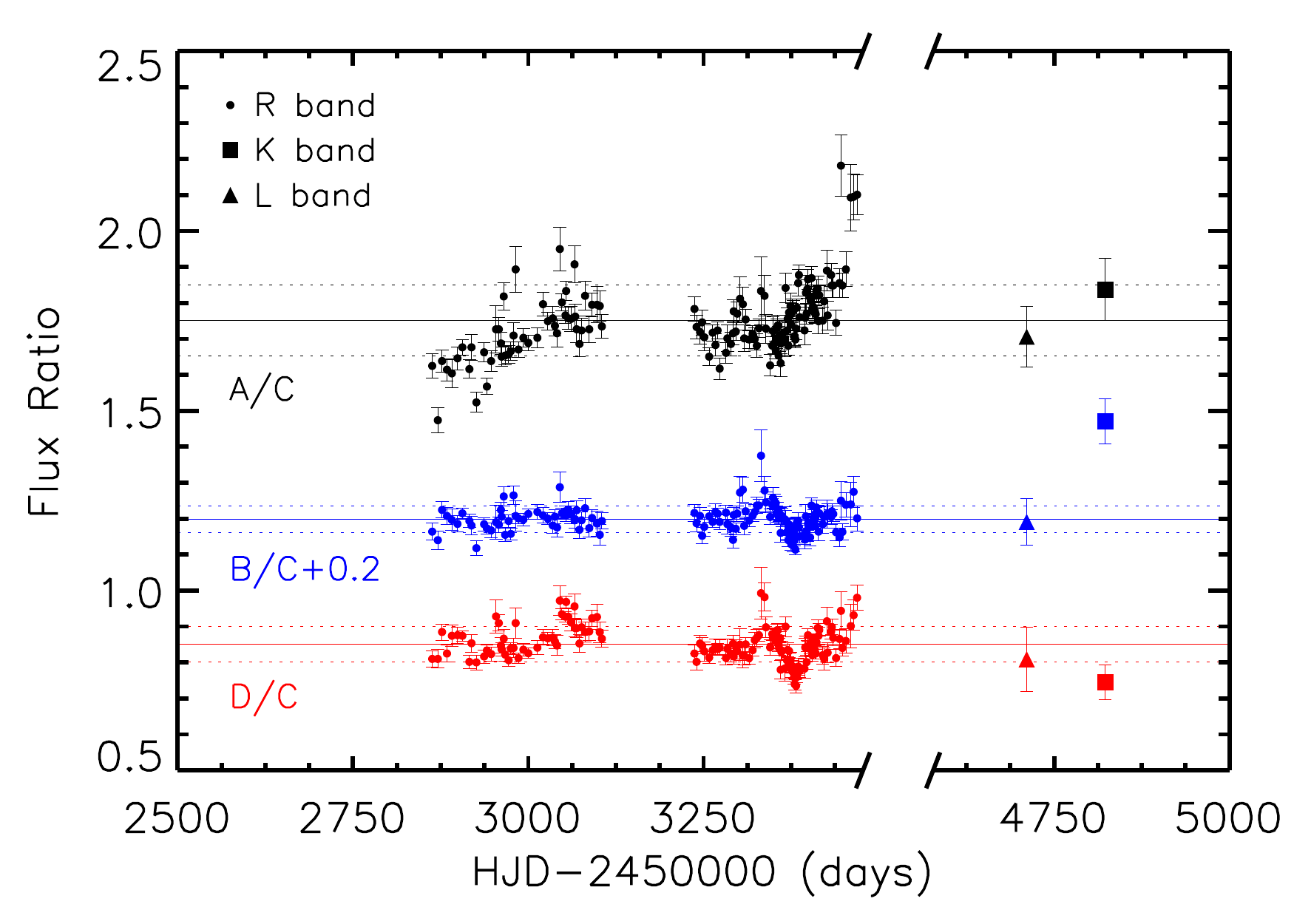}
\caption{
Flux ratios of images A, B, and D, relative to image C, as a function of observation epoch.  (Section \ref{sec:few-approach} explains why we take flux ratios relative to image C.)  The $B/C$ flux ratios are offset by 0.2 for visual clarity.  Small circular points show $R$-band monitoring from \citet{kochanek06a}, while square and triangular points show single-epoch $K$- and $L'$-band data from \citet{fadely11a}.  Solid and dotted lines indicate the mean and 68\% confidence ranges across all epochs in the $R$-band data.
}\label{fig:0435lightcurve}
\end{figure}

More care must be given to the photometric data.  Several datasets are available:  \citet{kochanek06a} present $R$-band monitoring, \citet{mosquera11a} present photometry in one broad-band and six narrow-band filters spanning $\sim$3500--8100 \AA, \citet{wisotzki03a} present optical integral field spectroscopy, and \citet{fadely11a} present photometry in the near-infrared $K$ and $L'$ bands.  Figure \ref{fig:0435lightcurve} shows the main dependences on time and wavelength using the optical monitoring of \citet{kochanek06a} and the NIR flux ratios of \citet{fadely11a}.  Three key features are seen in these data.  There is clear time variability in the $R$-band flux ratios.  All three $L'$-band flux ratios are consistent with the mean values of the $R$-band flux ratios.  Two of the $K$-band flux ratios are likewise consistent with the other wavelengths, but the $K$-band value of the $B/C$ flux ratio is a factor of $\sim$1.3 higher than the corresponding $R$- and $L'$-band values.  One other key result, from analysis of spectra by \citet{wisotzki03a}, is that there is no evidence of dust extinction in the lens galaxy.

The subtlety here is that the measured flux ratios may be affected by stellar microlensing, but we would prefer to omit microlensing from our analysis to the extent possible because it adds considerable computational expense and distracts from our focus on dark matter substructure.  Thus, we need to understand whether it is possible to account for or even eliminate microlensing from the flux ratio constraints.  One simple possibility is to broaden the errorbars so they encompass microlensing effects.  We do that by computing the standard deviation across all epochs in the $R$-band light curves from \citet{kochanek06a}.  This incorporates all microlensing and intrinsic variability that occurred during the 2-year span of the light curves.

It does not, however, account for microlensing effects with time scales longer than $\sim$2 yr.  To take the analysis one step further, we consider the multi-wavelength structure of the source quasar.  For a source redshift of $z_s = 1.689$, the $R$- and $K$-band observations probe rest-frame UV and optical wavelengths that are dominated by thermal emission from the hot quasar accretion disk, which is small enough ($\sim\!10^{15-17}$ cm; \citealt{morganc10a}) to be sensitive to microlensing.  By contrast, the $L'$-band observations (rest frame 1.4 $\mu$m) should contain emission from both the accretion disk and the surrounding dust torus \citep{rowanrobinson95a,nenkova08a}.  The relative contributions of the two components are not known exactly, but the dust torus probably accounts for 20--80\% of the flux \citep[e.g.,][]{wittkowski04a,honig08a}.  Since the dust torus should be large enough to be immune to microlensing, its contribution should cause the $L'$ flux ratios to have little or no variability from microlensing.  We therefore interpret the similarity between the $L'$- and $R$-band flux ratios as evidence that there is no significant long-term microlensing affecting the $R$-band light curves.

In other words, we can take either the $L'$-band measurements or the $R$-band measurements (with broadened errorbars) as microlensing-free estimates of the flux ratios.  In some sense the choice is not very important because the two sets of measurements are consistent with each other.  In practice, it is easier to work with the $R$-band data because at these wavelengths the source is much smaller than the Einstein radius of mass clumps larger than $\sim\!100\,M_\odot$, so we can effectively treat it as a simple point source in our study of lensing by dark matter substructure \citep[cf.][]{dobler06a}.

In the context of this analysis we still need to understand why the $K$-band measurement of the $B/C$ flux ratio differs from the other measurements.  From Figure \ref{fig:0435lightcurve}, the $B/C$ flux ratio must vary significantly with time and/or wavelength.  Dark matter subhalos can be ruled out as the cause because the agreement between the $L'$ and $R$ flux ratios suggests that there is little or no ``chromatic millilensing'' in HE0435 \citep[see][]{dobler06a}.  Microlensing may be a viable explanation, though, if the Einstein radii of stars in the lens galaxy are comparable to or larger than the size of the $K$-band source.  We examine this hypothesis carefully in Section \ref{sec:microK}, considering not only how microlensing might explain the $K$-band data but also whether it could have altered the $L'$-band data as well.  To jump ahead, we conclude that microlensing can indeed explain the $K$-band data without ruining our interpretation that the $L'$-band data provide good estimates of the microlensing-free flux ratios.

In our primary modeling we elect not to use measured time delays as constraints.  At the time of our analysis, \citet{kochanek06a} had published time delays based on two seasons of monitoring, but it was not clear how well the quasar and microlensing variability had been disentangled.  Indeed, \citet{blackburne10a} reported newer time delay estimates differed by 2--5$\sigma$ from the previous results.  After our analysis was complete, \citet{courbin10a} presented new data for HE0435 including refined astrometry from deconvolution of HST images and time delays from four additional years of $R$-band monitoring.  We compare the new time delays to our lens models in Section \ref{sec:delays}.  One valuable by-product of the analysis by \citet{courbin10a} is estimates of the $R$-band flux ratios after correcting for both microlensing and intrinsic variability in the source.  We note that these ``corrected'' $R$-band flux ratios match within $\sim\!1\sigma$ the mean $R$-band values used here. 

%=====================================================================
\section{Methodology}
\label{sec:method}
%=====================================================================

We use a Bayesian framework both to constrain model parameters and to assess the quality of different models.  We aim to compute the posterior probability distribution
\begin{eqnarray} \label{eqn:posterior}
  P(\theta|d,M)=\frac{P(d|\theta,M)\,P(\theta|M)}{P(d|M)}
\end{eqnarray}
where $d$ is the data which constrain the parameters $\theta$ for model $M$.\footnote{Note that $d$ and $\theta$ can be vectors, but we omit vector notation here for simplicity.}  We calculate the likelihood, $\mathcal{L}=P(d|\theta,M)$, from the $\chi^2$ goodness-of-fit: $\mathcal{L} \propto e^{-\chi^2/2}$.  Since we are only concerned with \emph{relative} posterior probabilities, we ignore the proportionality constant and set $\mathcal{L} = e^{-\chi^2/2}$.  In most cases we take the prior distribution, $P(\theta|M)$, to be uniform for the parameters listed in Table \ref{tab:parms}; one exception is discussed in Section \ref{sec:pop-approach}.

The denominator in eqn.~(\ref{eqn:posterior}) is the marginal likelihood of the model, also known as the Bayesian Evidence:
\begin{eqnarray}
  {\rm Evidence}(M) = P(d|M) = \int P(d|\theta,M)\,P(\theta|M)\ d\theta
\end{eqnarray}
In many astrophysical studies there is only one model being examined.  In that case the Bayesian evidence can be ignored, since the normalisation of the posterior does not affect confidence intervals for marginalised parameters.  If the evidence is not needed, an effective way to proceed is to sample from the posterior distribution using methods such as Monte Carlo Markov Chains (MCMC).

The evidence becomes crucial, though, when comparing different models.  Since the Bayesian evidence quantifies the overall probability of a particular model, it provides an objective way to compare models even if they have different numbers of parameters \citep{mackay03,gelman03}.  The ratio of the posterior probabilities for two models $M_1$ and $M_2$ is
\begin{eqnarray}
  \frac{P(M_2|d)}{P(M_1|d)} = \frac{P(d|M_2)}{P(d|M_1)}\,\frac{P(M_2)}{P(M_1)}
\end{eqnarray}
We assume equal prior probabilities for all models, $P(M_1)=P(M_2)$, so the ratio of posterior probabilities is just the ratio of evidences, which is called the Bayes factor.  While the principle is well established, the quantitative significance of Bayes factors is not completely clear cut, and various scales are employed to facilitate the interpretation.  The most common choice is the Jeffreys' scale \citep{jeffreys61}, which grades Bayes factors as shown in Table \ref{tab:jeffreys}.  In this work, we use the Jeffreys' scale as a guideline for judging our models, and we actually work with the differential log evidence, $\Delta\log_{10}({\rm Evidence}) = \log_{10}({\rm Bayes\ factor})$.

\begin{table}
\begin{center}
\begin{tabular}{cc}
\hline
$\Delta\log_{10}({\rm Evidence})$ & Significance \\
\hline
0--0.5 & Barely worth mentioning \\
0.5--1.0 & Substantial \\
1.0--1.5 & Strong \\
1.5--2.0 & Very strong \\
$>2.0$ & Decisive \\
\hline
\end{tabular}
\caption{
Jeffreys' scale \citep{jeffreys61} for grading the significance associated with different ranges of the (logarithmic) Bayes factor.
}\label{tab:jeffreys}
\end{center}
\end{table}

The practical challenge lies in integrating over all model parameters to compute the Bayesian evidence.  We perform the integration using the Nested Sampling algorithm \citep{skilling04a,skilling06a}, which provides marginalised parameter ranges in addition to the evidence.   As a computational tool, nested sampling has been used in a variety of astrophysical studies \citep[e.g.,][]{mukherjee06a,humphrey09a}, including gravitational lensing \citep[e.g.,][]{vegetti09b,barnabe09a}.

Roughly speaking, the idea of nested sampling is to execute many random draws from the parameter space according to the following scheme: at each step, the next point is drawn \emph{uniformly} from the prior distribution but limited to the region where the likelihood \emph{increases}.  Various procedures for doing this constrained sampling have been introduced \citep{mukherjee06a,shaw07a,feroz08a,feroz09a}; it is also possible to do nested sampling without such a strict one-way progression \citep{brewer09a}.  The volumes enclosed by different iso-likelihood surfaces may not be known exactly, but they can be estimated statistically, so the likelihood values and associated volumes can be combined to estimate the Bayesian evidence \citep[for details, see][]{skilling04a,skilling06a}.  Because nested sampling is intrinsically stochastic, there is some statistical uncertainty in the evidence value, but we can compute the uncertainty using the methods presented by \citet{keeton11a}.

In general we do not wish to place strong priors on our model parameters, so we have a large parameter volume to explore.  To alleviate the computational burden, we adopt a two-step approach to sampling.  First, we execute an MCMC sampling of the posterior using uniform priors defined in Table \ref{tab:parms}.  Details of our MCMC algorithm, techniques, and convergence criteria are discussed in Section 3.4 of \citet{fadely10a}.  We use the posterior from MCMC to construct narrower priors that encompass the 99.999\% CL parameter ranges (Table \ref{tab:parms}, dotted lines Figure \ref{fig:diagram}).  Using the narrower priors for nested sampling reduces the amount of time spent in regions of extremely low likelihood ($\chi^2>10^6$).  Tests with multivariate Gaussian distributions indicate that truncating such low-likelihood regions of the parameter space does not significantly alter estimates of the Bayesian evidence.

\begin{table}
\begin{center}
\begin{tabular}{ccc}
\hline
Parameter & MCMC prior & Nested Sampling prior \\
\hline
\multicolumn{3}{c}{Minimal, smooth model} \\
$\log_{10}(b_{\rm G1}/'')$ & $  -\infty\phantom{'''} : \phantom{''}\infty $ &$  \phantom{-}0.02\phantom{'''} : \phantom{-}0.13\phantom{''}  $ \\
$x_{\rm G1}$ & $ -\infty\phantom{'''} : \phantom{''} \infty $ &$  -0.003''\phantom{'} :\phantom{-}0.003''  $ \\
$y_{\rm G1}$ & $ -\infty  \phantom{'''} : \phantom{''} \infty $ &$  -0.003''\phantom{'} : \phantom{-}0.003''  $ \\
$e_{\rm c,G1}$ & $ -1.0  \phantom{'''} : \phantom{''}1.0 $ &$ -0.50\phantom{'''}  : \phantom{-} 0.50\phantom{''}  $ \\
$e_{\rm s,G1}$ & $ -1.0  \phantom{'''} : \phantom{''}1.0 $ &$ -0.50\phantom{'''}  :  \phantom{-}0.50\phantom{''}  $ \\
$\gamma_{\rm c}$ & $ -1.0  \phantom{'''} : \phantom{''}1.0 $ &$  -0.04\phantom{'''}  :  \phantom{-}0.06\phantom{''}  $ \\
$\gamma_{\rm s}$ & $ -1.0  \phantom{'''} : \phantom{''}1.0 $ &$  -0.03\phantom{'''} : \phantom{-}0.03\phantom{''}  $ \\
$s_{\rm G1}$ & $ 0.00''  \phantom{'} : \phantom{''} \infty $ &$  \phantom{-}0.00'' \phantom{'}:  \phantom{-}0.02''  $ \\
$\beta_{\rm G1}$ & $ -\infty  \phantom{'''} : \phantom{''} \infty $ &$ \phantom{-}0.95\phantom{'''}  : \phantom{-}1.60\phantom{''}  $ \\
$\log_{10}(b_{\rm G22}/'')$ & $ -1.7\phantom{''}  : \phantom{''} \infty $ &$  -1.70\phantom{'''} :  -0.12\phantom{''}  $ \\
$x_{\rm G22}$ & $ -\infty  \phantom{'''} : \phantom{''} \infty $ &$  \phantom{-}2.572''\phantom{'} :  \phantom{-}2.597''  $ \\
$y_{\rm G22}$ & $ -\infty  \phantom{'''} : \phantom{''} \infty $ &$  -3.625''\phantom{'}  :  -3.650'' $ \\
$e_{\rm c,G22}$ & $ -1.0  \phantom{'''} : \phantom{''}1.0 $ &$  -0.70\phantom{'''}  :  \phantom{-}0.70\phantom{''}  $ \\
$e_{\rm s,G22}$ & $ -1.0 \phantom{'''} : \phantom{''}1.0 $ &$ -0.70\phantom{'''} : \phantom{-}0.70\phantom{''}  $ \\
\hline
\multicolumn{3}{c}{Clump models} \\
$\log_{10}(b_{\rm A}/'')$ & $ -\infty \phantom{'''} : \phantom{''}\infty $ &$  -4.00\phantom{'''} : -1.00 \phantom{''} $ \\
$x_{\rm A}$ & $ -\infty\phantom{'''} : \phantom{''} \infty $ &$  -1.40''\phantom{'} : -0.70''  $ \\
$y_{\rm A}$ & $ -\infty  \phantom{'''} : \phantom{''} \infty $ &$  \phantom{-}0.40''\phantom{'} : \phantom{-}0.80''  $ \\
$\log_{10}(b_{\rm B}/'')$ & $ -\infty \phantom{'''} : \phantom{''}\infty $ &$  -4.00\phantom{'''} : -1.00 \phantom{''} $ \\
$x_{\rm B}$ & $ -\infty\phantom{'''} : \phantom{''} \infty $ &$   \phantom{-}0.20''\phantom{'} :\phantom{-}0.80''  $ \\
$y_{\rm B}$ & $ -\infty  \phantom{'''} : \phantom{''} \infty $ &$  \phantom{-}1.00''\phantom{'} : \phantom{-}1.50''  $ \\
$\log_{10}(b_{\rm D}/'')$ & $ -\infty \phantom{'''} : \phantom{''}\infty $ &$  -4.00\phantom{'''} : -1.00 \phantom{''} $ \\
$x_{\rm D}$ & $ -\infty\phantom{'''} : \phantom{''} \infty $ &$   -0.40''\phantom{'} :\phantom{-}0.50''  $ \\
$y_{\rm D}$ & $ -\infty  \phantom{'''} : \phantom{''} \infty $ &$  \phantom{-}1.00''\phantom{'} : \phantom{-}1.50''  $ \\
\hline
\end{tabular}
\caption{
Model parameters and priors for our smooth and clump models discussed in Sections \ref{sec:smooth} and \ref{sec:few}, respectively.  We adopt uniform priors within the specified intervals, with the exception of $e_{\rm c,G22}$ and $e_{\rm s,G22}$ for certain models as discussed in Section \ref{sec:pop-approach}.  Note that $e_c = e \cos 2\theta_e$ and $e_s = e \sin 2\theta_e$ are the quasi-Cartesian components of the ellipticity $e=1-q$.
}\label{tab:parms}
\end{center}
\end{table}

%=====================================================================
\section{Smooth Models}
\label{sec:smooth}
%=====================================================================

As a first step we examine how well a smooth mass distribution (without substructure) can fit the observed image positions and flux ratios for HE0435.  Following \citet{kochanek06a}, we adopt a ``minimal'' lens model in which the main lens galaxy (G1) has an ellipsoidal mass distribution with a softened power law density profile,
\begin{eqnarray}
\kappa(\xi)=\frac{1}{2}\frac{b_{\rm G1}^{2-\beta}}{(s^2+\xi^2)^{1-\beta/2}}
\end{eqnarray}
where $s$ is the core radius, $\xi=\sqrt{x^2+y^2/q^2}$ is the ellipse coordinate (in the major axis frame), and $q \le 1$ is the projected axis ratio.  The power law index is defined such that the mass enclosed within radius $R$ scales as $M(R) \propto R^\beta$, so an isothermal profile has $\beta=1$, while a steeper (shallower) profile has $\beta<1$ ($\beta>1$).  Note that for a pure power law profile, the corresponding 3D density profile is $\rho \propto r^{\beta-3}$.  The normalisation parameter $b$ has dimensions of length, and for a pure power law $b$ is directly proportional to the Einstein radius: $\Rein = b F(\beta,q)$.  The proportionality factor $F$ has a complicated form in terms of special functions, but if we make a Taylor expansion in the ellipticity $e=1-q$ we can write
\begin{eqnarray}
  F = \beta^{\frac{1}{\beta-2}}
  \left[1-\frac{e}{2}-\frac{2+\beta}{16}\,e^2-\frac{2+\beta}{32}\,e^3+{\cal O}(e^4)\right] .
\end{eqnarray}
We vary all model parameters for the main lens, including $\beta$.  \citet{kochanek06a} found that HE0435 has a density profile that is shallower than isothermal, which leads to a rising rotation curve.  Varying $\beta$ is important if we are to account for a wide range of possible mass distributions.

Our minimal model also includes the effects of the environment of HE0435.  The neighboring galaxy G22 is close enough (only $4.4''$ from G1) that it needs to be included explicitly.  As in previous studies, we assume that G22 lies at the same redshift as G1.  \citet{kochanek06a} modeled G22 as a singular isothermal sphere and found a best-fit Einstein radius of $0.22''$, so the galaxy should provide negligible surface mass density at the location of HE0435, and an isothermal profile should be adequate.  We do, however, generalise by letting G22 be elliptical and use a singular isothermal ellipsoid model.  We also add an independent external shear to account for tidal effects from the group of galaxies surrounding the lens \citep{morgann05a,wongkc11a}.

\begin{figure*}
\centering
\includegraphics[clip=true, trim=2.2cm 12.6cm 2.1cm 3.cm,width=8cm]{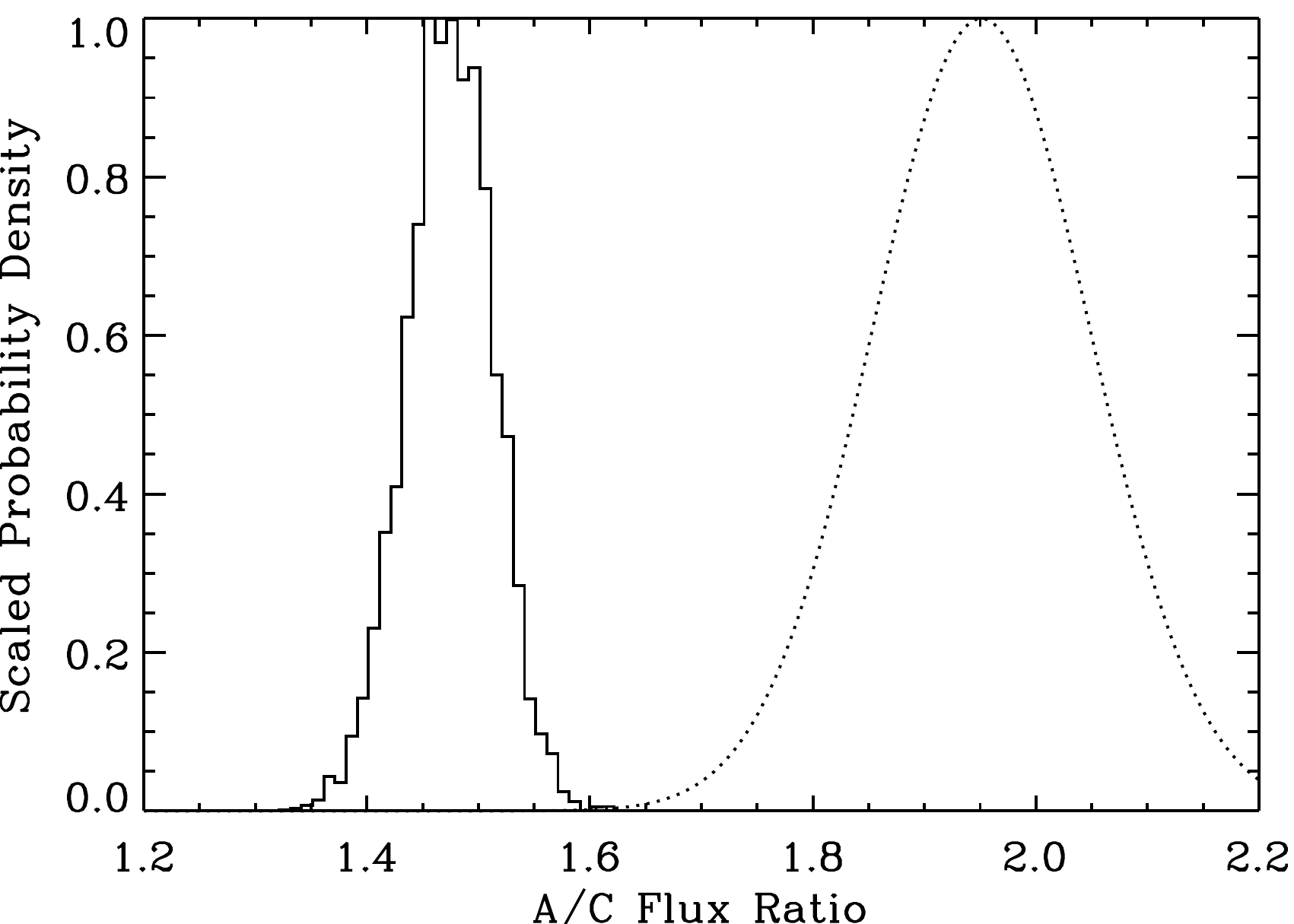}
\includegraphics[clip=true, trim=2.2cm 12.6cm 2.1cm 3.cm,width=8cm]{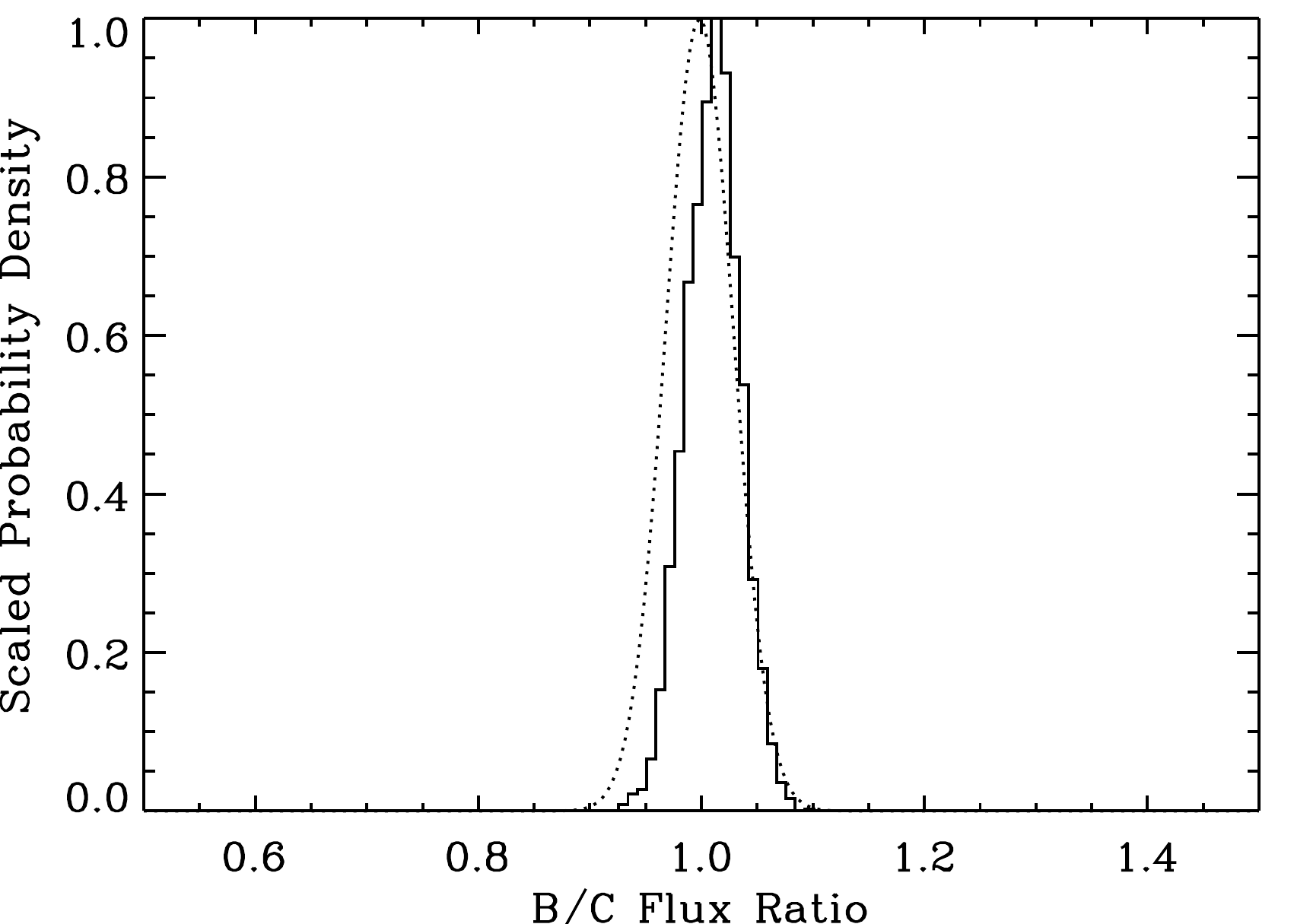}
\includegraphics[clip=true, trim=2.2cm 12.6cm 2.1cm 3.cm,width=8cm]{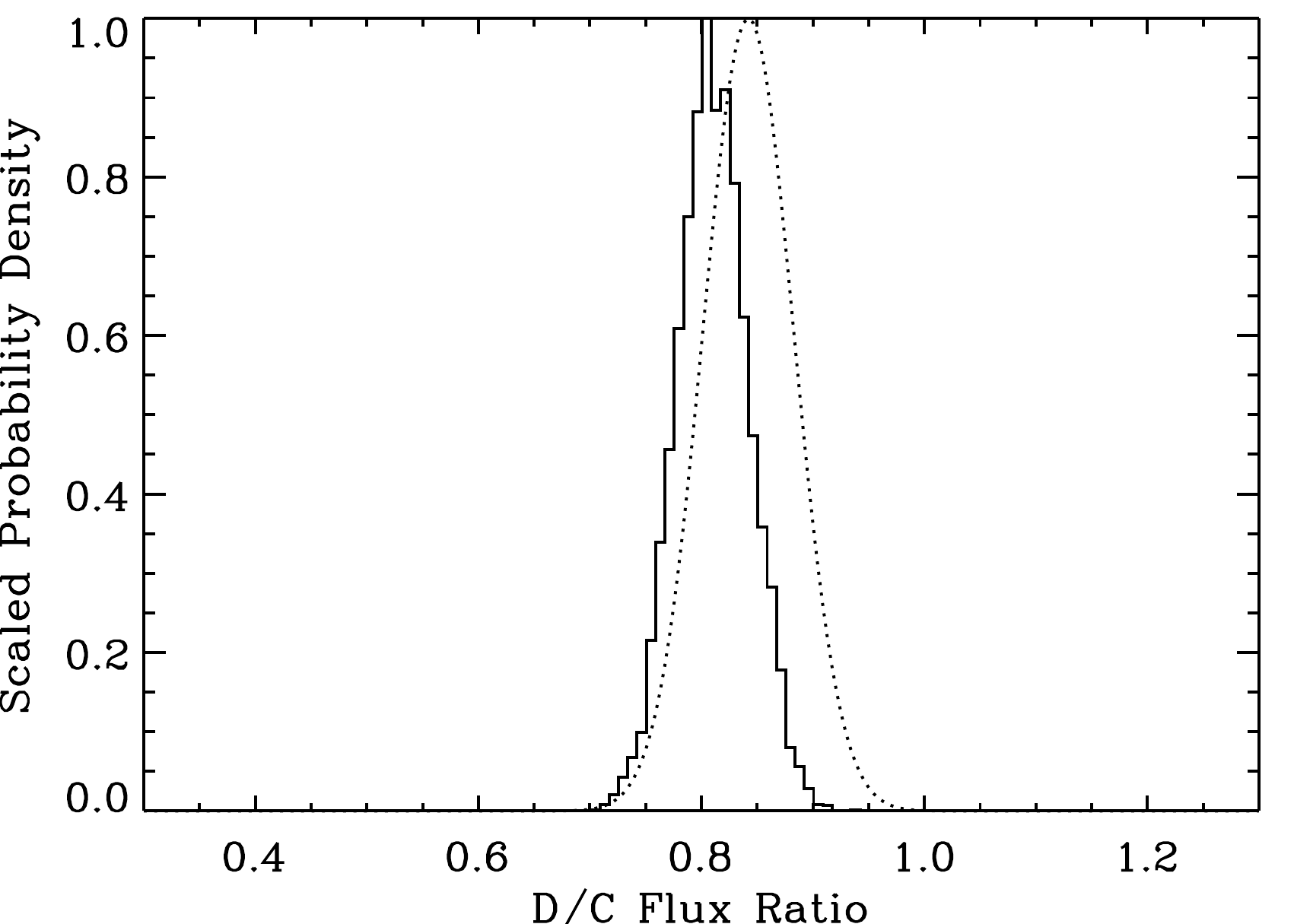}
\caption{
Solid lines show marginalised probability distributions for the $R$-band flux ratios inferred from our smooth, minimal mass model.  Dotted lines show the likelihood functions for the observed $R$-band distributions (Gaussians with the measured mean and variance).  All curves are normalised to have a peak of unity.  The minimal model can match the observed $B/C$ and $D/C$ flux ratios, but cannot match $A/C$.
}\label{fig:fluxsmooth}
\end{figure*}

We use an updated version of the public \texttt{lensmodel} code \citep{Keeton01a} both to find the best-fit smooth model and to run nested sampling.  The resulting Bayesian evidence is reported in Table \ref{tab:results} below.  The best-fit model has $\chi^2=24.6$ for $N_{\rm dof}=-1$.  Since the model is formally underconstrained, finding $\chi^2 \ne 0$ indicates that the model is not sufficiently flexible---it lacks some key freedom.  To diagnose the failure, we show in Figure \ref{fig:fluxsmooth} the distributions of flux ratios predicted by the smooth model, compared with the observed values.\footnote{Strictly speaking, if the fluxes are normally distributed then the flux ratios follow Lorentzian distributions.  For practical purposes, though, we treat the flux ratios using Gaussian distributions.}  The smooth model is unable to account for the observed $A/C$ flux ratio at high confidence.  This constitutes a clear ``flux ratio anomaly'' of the sort that has been seen in other lenses \citep[e.g.,][]{maos98a,bradac02a,metcalf02a,keeton03a,keeton05a} and interpreted as evidence for dark matter substructure \citep[e.g.,][]{metcalf01a,dalal02a}.

%=====================================================================
\section{Few-Clump Models}
\label{sec:few}
%=====================================================================

Motivated by the anomaly, we hypothesise that HE0435 contains substructure and examine how well we can test that hypothesis and constrain the properties of the substructure.  In this section we consider whether it is possible to explain the data with just one or a few clumps that (presumably) lie near the lensed images.  In Section \ref{sec:pop} we study full populations of substructure.

%=====================================================================
\subsection{Approach}
\label{sec:few-approach}

We first need to reflect on where we might expect mass clump(s) to be.  Since the observed $A/C$ flux ratio is higher than predicted by smooth models, we need substructure to increase the predicted ratio, which means making image A brighter or C fainter.  In HE0435, A and C are both positive-parity images.  Substructure tends to make positive-parity images brighter and negative-parity images fainter \citep{schechter02a,keeton03b}, so we expect to need a clump near image A.  We do not place a clump near image C, because that would tend to make C brighter and exacerbate the flux ratio anomaly.  (This is the reason we have chosen to compute flux ratios relative to image C.)

We imagine that there might be additional clumps near images B and/or D.  To be specific, we consider a model with one clump near image A, a model with two clumps near A and B, a different model with two clumps near A and D, and a model with three clumps near A, B, and D.  For simplicity, we label these four models A, AB, AD, and ABD, respectively.  The models clearly have different numbers of parameters, but the Bayesian analysis can provide an objective ranking of the models (through the evidence) along with constraints on the clump positions and masses (through parameter marginalisation).

We model the clumps with a spherical pseudo-Jaffe profile, which has a three-dimensional density $\rho(r) \propto 1/r^2(a^2+r^2)$ that translates into a two-dimensional surface mass density of the form
\begin{eqnarray} \label{eqn:clump}
  \kappa(r)=\frac{b_{\rm clump}}{2}\left [\frac{1}{r}-\frac{1}{\sqrt{a^2+r^2}}\right ]
\end{eqnarray}
where $b_{\rm clump}$ sets the mass scale while $a$ represents a truncation radius.  The total mass for this model is $M_{\rm total}=\pi\Sigma_{\rm crit}b_{\rm clump}a$.  Pseudo-Jaffe clumps are efficient lenses because they have a steep, isothermal slope inside the truncation radius; clumps with shallower profiles (e.g., NFW) are less efficient lenses and could therefore lead to different model results.  We select the pseudo-Jaffe model because it includes the effects of tidal truncation, and its role in earlier studies \citep[e.g.,][]{dalal02a,vegetti10a} facilitates the comparison of previous results with our results using new methodology.  We defer a systematic study of clump density profiles to follow-up work.  Following \citet{dalal02a}, we set $a=\sqrt{\langle b_{\rm G1}\rangle b_{\rm clump,max}}$ to account for tidal truncation of a pseudo-Jaffe profile by the parent halo in an approximate but reasonable way.  Here $\langle b_{\rm G1}\rangle$ is the average mass normalisation of G1 and $b_{\rm clump,max}$ is the maximum mass scale parameter for the clump.  For HE0435, this works out to be $a=0.367''$.

Table \ref{tab:parms} provides a complete list of the parameters for our few-clump models.  We vary all of the parameters using nested sampling to obtain both evidence values and parameter constraints.  We then use the constrained clump parameters to compute the clump masses.  We can compute the total mass of the pseudo-Jaffe model, but we expect the quantity that is more relevant for lensing (especially flux ratios) to be the mass within the Einstein radius.  We quote both but give particular attention to the mass within the Einstein radius.

%=====================================================================
\subsection{Results}
\label{sec:few-results}

\begin{figure}
\centering
\includegraphics[clip=true, trim=2.2cm 12.6cm 2.1cm 3.cm,width=8cm]{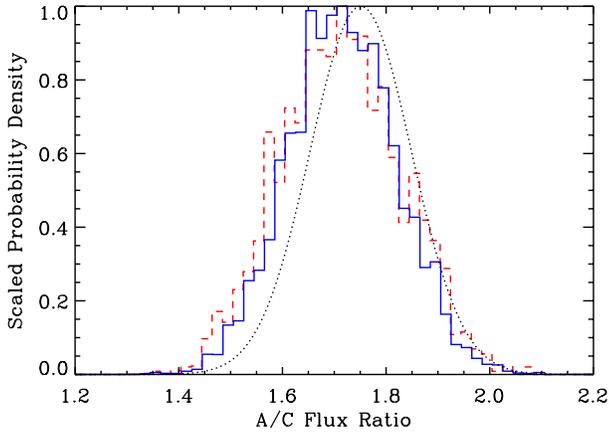}
\caption{
Similar to the top left panel of Figure \ref{fig:fluxsmooth}, but for our model with a mass clump near image A (solid, blue) and our model with three clumps near images A, B, and D (dashed, red).  Adding a clump near image A clearly brings the models into agreement with the data.  Adding clumps near images B and D has little effect on the predicted $A/C$ flux ratio, indicating that constraints on clump A are fairly independent of the presence of clumps near other images.
}\label{fig:fluxclump}
\end{figure}

We first add a single clump near image A to our macromodel.  Table \ref{tab:results} gives the Bayesian evidence along with constraints on the clump parameters.  Figure \ref{fig:fluxclump} shows that adding the clump does much to alleviate the discrepancy between the predicted and observed flux ratios: the predicted value is now $A/C = 1.72^{+0.11}_{-0.10}$.  In fact, the model is able to reproduce the data perfectly, with a best-fit value of $\chi^2=0$.  In some sense this is not surprising because the model is underconstrained with $N_{\rm dof}=-4$, but we saw before that being underconstrained does not guarantee a perfect fit.  Apparently the clump provides some important freedom that was not present in the smooth, minimal model.

\begin{figure}
\centering
\includegraphics[clip=true, trim=2.2cm 12.5cm 1.9cm 2.cm,width=8cm]{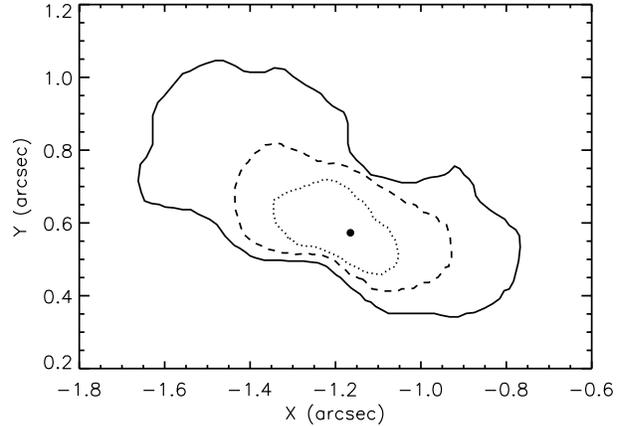}
\caption{
95\% confidence constraints on the position of the clump near image A (marginalised over all model parameters).  The circle indicates the position of image A.  Dotted, dashed, and solid contours show results for clumps with masses less than $10^6,10^7,$ and $10^8\,M_\odot$, respectively.  More massive clumps can lie farther from image A and still reproduce the observed flux ratio.
}\label{fig:clumppos}
\end{figure}

Figure \ref{fig:clumppos} shows constraints on the position of the clump near image A for different mass ranges.  We find that the clump position and mass are degenerate in the sense that a more massive clump can lie farther from the image and still reproduce the observed flux ratio.  This is familiar from previous studies \citep[e.g.,][]{dalal02a,keeton09b}, and not surprising because, heuristically, flux perturbations are driven by shear perturbations of the form $\delta\gamma \propto M/d^2$, where $d$ is the distance of the clump from the image \citep[see][]{maos98a}.  In principle, a star placed close to the image can produce the same magnification as a more massive clump placed farther away (provided the source is sufficiently small).  Adding position constraints can break the degeneracy, though.  Position perturbations are driven by deflection perturbations of the form $\delta\alpha \propto M/d$ \citep{chenj07a,keeton09b}.  Thus, if both the position and flux are affected by a clump, the different scalings make it possible to constrain the clump mass.  On the one hand, a very low-mass clump is simply unable to affect the image position, regardless of its location; on the other hand, a high-mass clump may disturb the image position too much, and may affect other images as well.  In HE0435 these effects let us find bounds on the mass of clump A: $\log_{10}(\Mein^A)=7.65^{+0.87}_{-0.84}$ (or $\log_{10}(M_{\rm total}^A) = 9.31^{+0.44}_{-0.42}$).\footnote{As noted in Section \ref{sec:few-approach}, we emphasise the mass within the Einstein radius because we expect it to be the more robust quantity.  The total mass is more sensitive to the clump profile and truncation radius.}  We conclude that the clump is well constrained by the combination of both flux and position data.

We next consider models with more than one clump near the images.  We keep the clump near image A because it seems essential, but try placing additional clumps near B and/or D.  Table \ref{tab:results} gives the evidence values and parameter constraints for the various models.  Figure \ref{fig:fluxclump} shows that adding more clumps does not significantly alter the predicted $A/C$ flux ratio.  This, in turn, implies that additional clumps have little effect on the inferred properties of clump A (see Table \ref{tab:results}).

\begin{table*}
\begin{tabular}{crrrr}
\hline
 Model:  & +A & +AB & +AD & +ABD \\
\hline
 $\Delta\log_{10}({\rm Evidence})$ & $3.83\pm0.16$ & $4.46\pm0.16$ & $3.90\pm0.18$ & $4.35\pm0.18$ \\
 $\log_{10}(\Mein^A)$& $7.46^{+0.57}_{-0.69}$ & $7.65^{+0.87}_{-0.84}$ & $7.30^{+0.58}_{-0.61}$ & $7.47^{+0.66}_{-0.68}$ \\
 $x_A$ & $-1.05^{+0.13}_{-0.17}$ & $-1.13^{+0.08}_{-0.13}$ & $-1.13^{+0.08}_{-0.13}$ & $-1.07^{+0.09}_{-0.11}$ \\
 $y_A$  & $0.51^{+0.07}_{-0.06}$ & $0.56^{+0.05}_{-0.06}$ & $0.56^{+0.05}_{-0.06}$ & $0.55^{+0.05}_{-0.05}$ \\
 $\log_{10}(\Mein^B)$  & $-$ & $6.55^{+1.01}_{-1.51}$ & $-$ & $6.14^{+1.32}_{-1.87}$ \\
 $x_B$  & $-$ & $0.15^{+0.24}_{-0.29}$ & $-$ & $0.45^{+0.14}_{-0.13}$ \\
 $y_B$  & $-$ & $1.04^{+0.12}_{-0.12}$ & $-$ & $1.31^{+0.10}_{-0.11}$ \\
 $\log_{10}(\Mein^D)$  & $-$ & $-$ & $5.80^{+1.52}_{-1.72}$ & $5.87^{+1.49}_{-1.80}$ \\
 $x_D$  & $-$ & $-$ & $0.15^{+0.19}_{-0.30}$ & $0.13^{+0.11}_{-0.22}$ \\
 $y_D$  & $-$ & $-$ & $-1.04^{+0.12}_{-0.13}$ & $-1.05^{+0.10}_{-0.11}$ \\
\hline
\end{tabular}
\caption{
Marginalised clump parameters and evidence values for models which add the specified clump(s) to our minimal, smooth model.  We quote differential log evidence values relative to the smooth model to facilitate model comparison (see Section \ref{sec:method} and Table \ref{tab:jeffreys}).  
Positions are given in arcsec.  The clump mass is the mass within the Einstein radius, in units of $h_{70}^{-1} M_\odot$.
}\label{tab:results}
\end{table*}

We can assess the relative probabilities of the various models using the Bayesian evidence values in Table \ref{tab:results}.  The models with clumps all have evidence values that are at least three orders of magnitude greater than our minimal smooth model.  Clearly, the data strongly prefer models with at least one clump near the lensed images.  According to the Jeffreys' scale (Table \ref{tab:jeffreys}), the case for at least one clump is decisive for the range of models considered here.  Examining the evidences in detail, we see that model AB has the highest evidence, with a value that is 0.63 dex higher than for the single-clump model.  Formally, this provides ``substantial'' evidence for a second clump near image B with mass $\log_{10}(\Mein^B)=6.55^{+1.01}_{-1.51}$ (or $\log_{10}(M_{\rm total}^B) = 8.76^{+0.50}_{-0.77}$).  Since the evidence values carry uncertainties of 0.16 dex, the case for clump B is intriguing but far from decisive.

It is striking to see that adding a clump near image D has essentially no effect on the Bayesian evidence: the evidences for the A and AD models are indistinguishable (given the uncertainties), and likewise for the AB and ABD models.  Apparently the parameters associated with clump D do not significantly improve the models' ability to reproduce the data, so they produce little change to the evidence.  This is an example of Occam's Razor in action.

\begin{figure*}
\centering
\includegraphics[clip=true, trim=1.25cm 12.2cm 2.0cm 3.1cm,width=8.5cm]{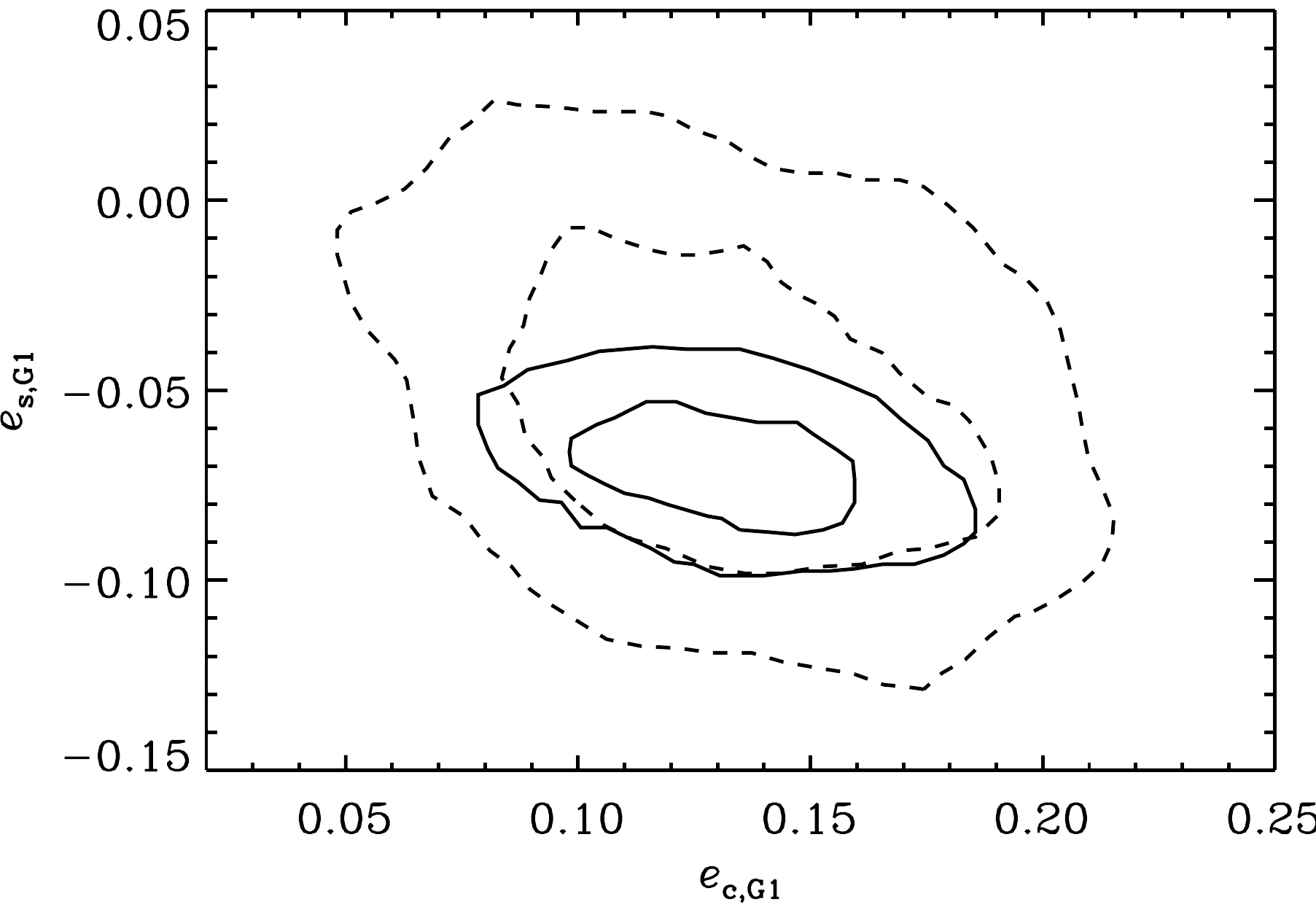}
\includegraphics[clip=true, trim=1.3cm 12.2cm 2.0cm 3.1cm,width=8.5cm]{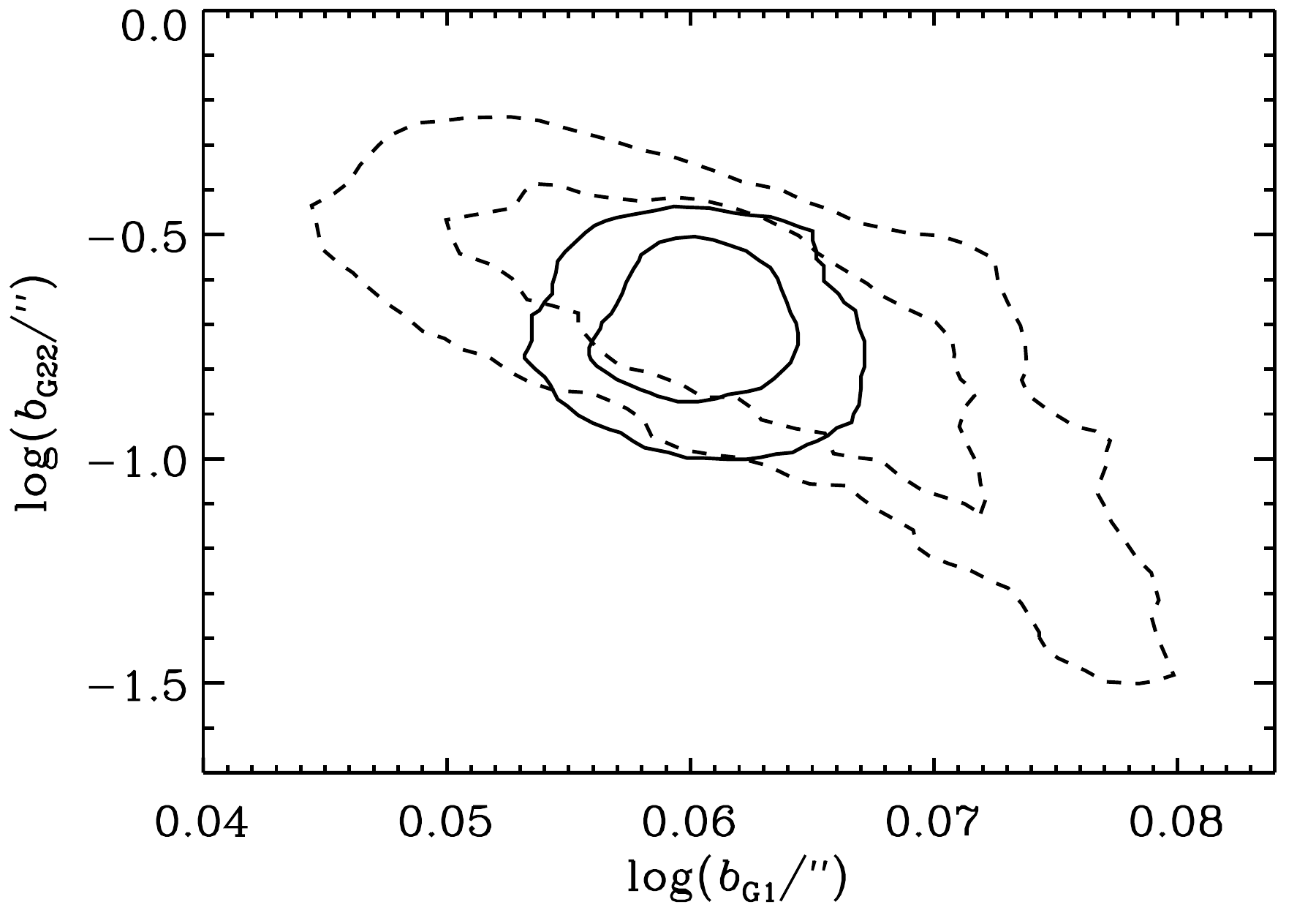}
\caption{
Joint posterior probability distributions for two pairs of parameters: the quasi-Cartesian components of the ellipticity of G1 ($e_c = e \cos 2\theta_e$ and $e_s = e \sin 2\theta_e$, left panel), and the mass normalisations of the lens G1 and its neighbor G22 (right panel).  The solid curves show the 68\% and 95\% confidence contours from our minimal, smooth model, while the dashed curves shows results from our model with a clump near image A.  Adding substructure can broaden the parameter distributions (e.g., left panel) and also introduce covariances (e.g., right panel).
}\label{fig:0435smoothparms}
\end{figure*}

It is interesting to study how the addition of clumps affects the parameters of the smooth mass distribution.  Figure \ref{fig:0435smoothparms} shows joint posterior probability distributions for several key parameters, before and after adding a clump near image A.  In general, adding the clump broadens the distributions, which is not surprising because of the increased flexibility afforded by the clump.  For some parameters the posterior also develops structure indicative of covariances in the 14-dimensional parameter space (see the right-hand panel of Fig.~\ref{fig:0435smoothparms}).  Even so, the median values of the distributions are not significantly altered, typically shifting within the 68\% confidence intervals of the no-clump model.

One notable parameter is the density slope of the main lens, $\beta$.  Using the quasar image positions and (estimated) time delays, the Einstein ring image of the quasar host galaxy, and a prior on $H_0$, \citet{kochanek06a} found the slope to be shallower than isothermal, corresponding to $\beta>1.0$ in our models.  We obtain $\beta=1.19^{+0.13}_{-0.13}$ and $1.19^{+0.17}_{-0.15}$ for our models without and with clump A, respectively.  Thus, we find evidence for a shallow density profile independent of time delay constraints, and that result is not affected by the presence of substructure.

The mass constraints we find on substructure in HE0435 are a first for quasar lenses.  Previous work in the radio and mid-infrared has provided good evidence for substructure but has not necessarily yielded upper and lower bounds on clump masses \citep[e.g.,][]{chibam05a,minezaki09a}.  Presumably the constraints from flux ratios and image positions in some lenses are not (yet) strong enough to determine clump masses.  More recently, studies of the lens systems B2045+265 \citep{mckean07a}, MG 2016+112 \citep{more09a}, and H1413+117 \citep{macleodc09a} have been able to place specific constraints on clump masses.  In all of those cases, however, the substructure was linked to a luminous satellite whose position could be constrained from direct observations.  Fixing the position breaks the position-mass degeneracy that is inherent in flux perturbations, and thus can lead to very good constraints on mass ($\sigma_M \sim 0.1$--0.3 dex).  Our results for HE0435 are novel in two ways.  First, we have been able to constrain clump positions and masses from lens data alone, with no direct observations of the clump(s).  This shows that it is possible to constrain clumps even if they are invisible.  Second, the masses we have found for clumps A and B are among the smallest found in any lens system to date.\footnote{As noted in Section \ref{sec:few-approach}, the model results---including the inferred clump masses---may depend on the choice of clump density profile.  However, our use of the same clump profile as in previous work means that it is fair to compare clump masses from different studies.}

While we believe these conclusions to be new and interesting, we do offer one cautionary note.  Our clump constraints have been derived in the context of a fairly simple macromodel.  Adding more flexibility to the macromodel, such as higher-order multipole modes or pixellated potential perturbations \citep[e.g.,][]{evans03,yoo05,yoo06,blandford01,koopmans05}, would presumably alter the inferred clump constraints (e.g., broaden the mass uncertainties).  However, \citet{congdon05a} found that such features alone cannot explain flux ratio anomalies, and \citet{yoo05,yoo06} found that elliptical symmetry seems to be a reasonable assumption for lens galaxies.  We therefore expect that adding reasonable flexibility might weaken but not eliminate the evidence for substructure in HE0435.

%=====================================================================
\section{Substructure Population Models}
\label{sec:pop}
%=====================================================================

So far we have concentrated on a few individual clumps near the lensed images of HE0435.  Those clumps are presumably not the only examples of substructure in the system, but rather special representatives of a larger population (special in that they lie near an image).  In this section we aim to constrain a full substructure population of the sort predicted by CDM \citep[e.g.,][]{diemand07a, springel08a}.

%=====================================================================
\subsection{Model}
\label{sec:pop-model}

We assume the clump population is characterised by a power law mass function of the form $dN/dM \propto M^\alpha$ with $\alpha=-1.9$ \citep{diemand07a, springel08a} and fixed lower and upper total mass thresholds of $M_{\rm total}=10^{7}M_\odot$ and $10^{10} M_\odot$, respectively.  We assume the clump positions are drawn from a uniform spatial distribution out to $10''$ from the centre of the lens.  While realistic substructure may not be spatially uniform \citep[e.g.,][]{zentner05,springel08a,nierenberg11a}, using a uniform distribution facilitates comparison with previous work \citep[e.g.,][]{dalal02a}.  Moreover, the choice of spatial distribution should not be terribly important for our results, because we focus on observables (image positions and flux ratios) that are mainly sensitive to clumps in the vicinity of the Einstein radius \citep[e.g.,][]{rozo06,keeton09b}.

We characterise the abundance of substructure using the mean convergence, $\kappa_s = \Sigma_s/\Sigma_{\rm crit}$, where $\Sigma_s$ is the mean surface mass density in substructure (averaged over many realisations), and $\Sigma_{\rm crit}$ is the critical surface mass density for lensing.  For lensing purposes it is convenient to work with the scaled clump mass, $m=M_{\rm total}/\Sigma_{\rm crit}$, which has units of angular area.  If the number density of clumps per unit mass of $dn/dm$, then the substructure convergence is
\begin{eqnarray} \label{eqn:kappas}
  \kappa_s=\int m\,\frac{dn}{dm}\ dm. 
\end{eqnarray}
Before undertaking extensive simulations, we would like to see if we can use the clumps inferred so far to estimate the properties of the larger population.  While such an estimate must be taken with a grain of salt, it may help guide our exploration of parameter space.  In Appendix A we present a toy model for the probability distribution $P(\kappa_s)$ based on the idea that a clump population should have one clump in a location that leads to a strong flux perturbation in image A.  That analysis leads to an estimate of $\kappa_s = 0.025^{+0.074}_{-0.022}$ (95\% CL).  Therefore, we consider the values of $ \kappa_s=[0.00022$, 0.00046, 0.001, 0.0022, 0.0046, 0.01, 0.022, 0.046, $0.10]$ in our simulations.

For technical reasons, the clumps used in our population models differ from those used in our few-clump models in two small ways.  First, instead of a smoothly truncated pseudo-Jaffe profile we use a sharply truncated isothermal profile whose density is $\rho \propto r^{-2}$ inside the truncation and zero outside.  Second, the truncation radius is not fixed but scales with clump mass such that the ratio of the truncation radius to the Einstein radius is fixed to a ratio of $\approx$60.  (When we use a wide range of clump masses, it seems more sensible to have the truncation radius scale with mass than to use a some fixed value.)  While in detail these differences may influence the distributions of deflections and magnifications produced by clumps, they should not significantly affect our results.

%=====================================================================
\subsection{Approach}
\label{sec:pop-approach}

The complete set of parameters for this analysis includes the smooth model parameters (denoted by $\vect{\theta}$), the substructure convergence ($\kappa_s$), and the positions and masses of individual clumps (denoted by $\vect{c}$).  The quantity we seek is the posterior distribution for $\kappa_s$ after marginalising over $\vect{\theta}$ and $\vect{c}$,
\begin{eqnarray} \label{eqn:clumppost1}
  P(\kappa_s|\vect{d}) = Z_{\rm tot}^{-1} \int \mathcal{L}(\vect{d}|\vect{c},\vect{\theta})\,P(\vect{c}|\kappa_s)\,P(\vect{\theta})\ d\vect{c}\ d\vect{\theta}
\end{eqnarray}
The likelihood $\mathcal{L}$ depends explicitly on the clump positions and masses and the smooth model parameters; $\kappa_s$ enters only implicitly, through the number of clumps, so we have written it in the priors $P(\vect{c}|\kappa_s)$.  The latter factor also includes the priors on the clump positions and masses described above.  Finally, $P(\vect{\theta})$ indicates priors on the smooth model parameters (see Table \ref{tab:parms}), and $Z_{\rm tot}$ is a normalisation factor.

\begin{figure*}
\centering
\includegraphics[clip=true, trim=1.9cm 12.4cm 2.1cm 3.cm,width=8cm]{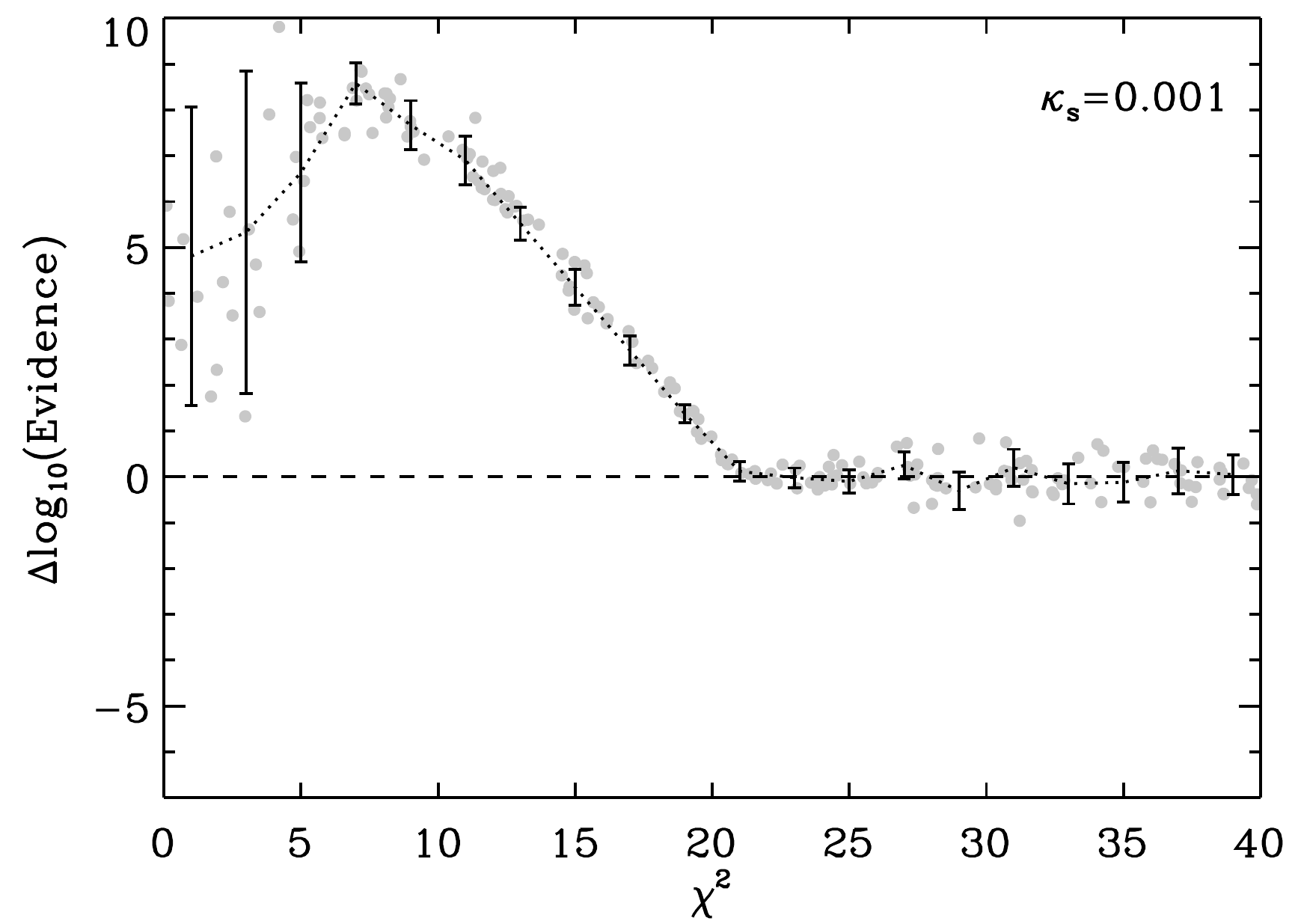}
\includegraphics[clip=true, trim=1.9cm 12.4cm 2.1cm 3.cm,width=8cm]{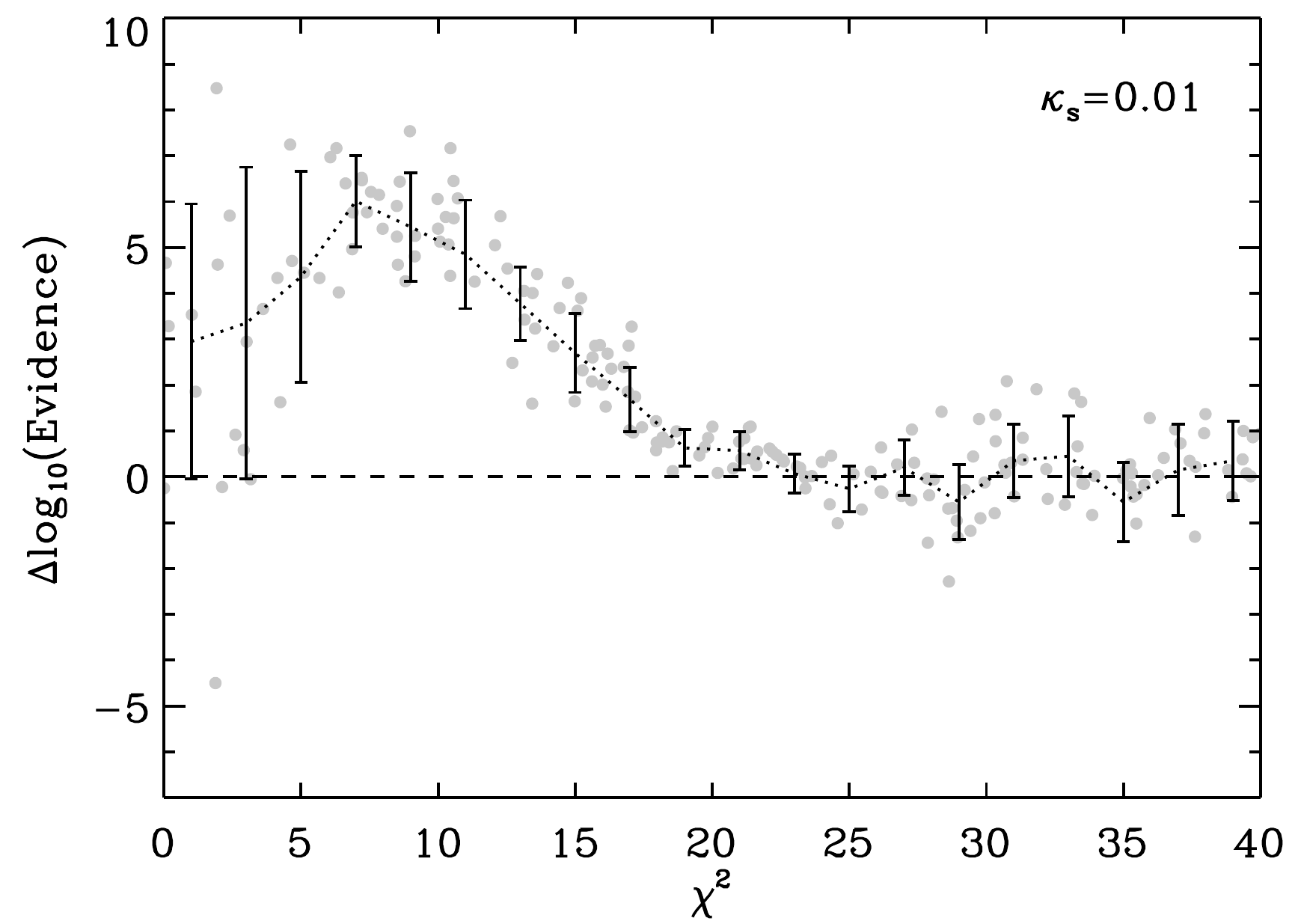}
\includegraphics[clip=true, trim=1.9cm 12.4cm 2.1cm 3.cm,width=8cm]{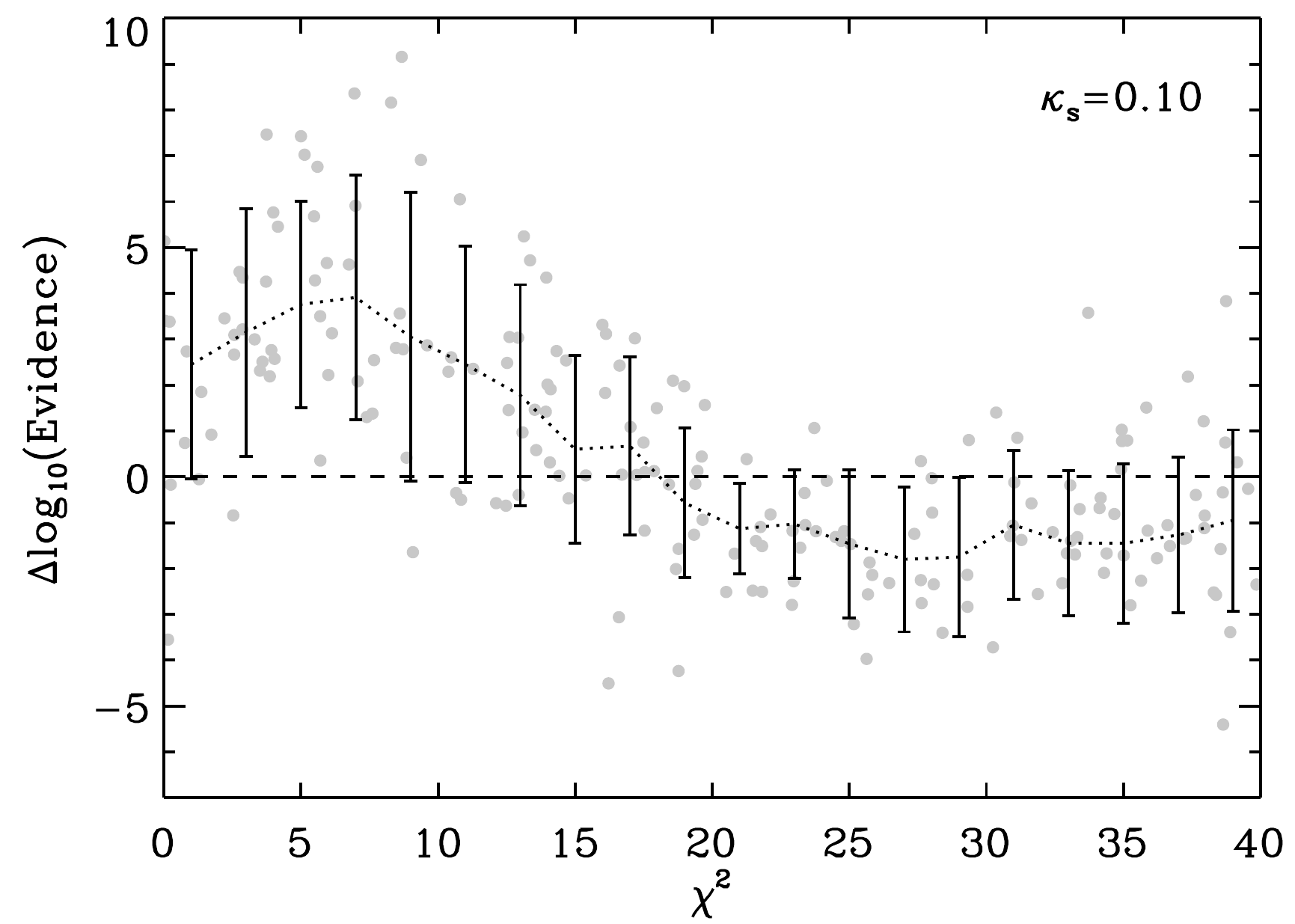}
\caption{
Best-fit $\chi^2$ values and evidences for a subset of our simulations (gray points).  We actually plot the differential log evidence relative to the smooth, minimal model.  The three panels show results for $\kappa_s = 0.001$, 0.01, and 0.10.  In all cases, the best-fit $\chi^2$ is a poor predictor of the evidence for $\chi^2 \la 6.5$.  For small values of $\kappa_s$, the evidence is tightly correlated with $\chi^2$ for $\chi^2 \ga 6.5$, and it is consistent with the smooth model value for $\chi^2 \ga 21$.  As $\kappa_s$ increases, so does the scatter in evidence values.  Overplotted are the mean and scatter in $\Delta\log_{10}({\rm Evidence})$ for $\chi^2$ bins, which are used to convert from $\chi^2$ to evidence for additional realisations (see text).
}\label{fig:chievid}
\end{figure*}

Formally the integrand in eqn.~(\ref{eqn:clumppost1}) may have hundreds or thousands of dimensions, depending on the number of clumps, so we cannot evaluate it directly.  Instead, we use Monte Carlo techniques.  Let $\vect{c}_j$ denote one particular realisation of the clump population.  Suppose we generate $N_c$ realisations for a particular value of $\kappa_s$.  Then heuristically we can let
\begin{eqnarray}
  \int f(\vect{c})\,P(\vect{c}|\kappa_s)\ d\vect{c} \ \to\ \frac{1}{N_c} \sum_{j=1}^{N_c} f(\vect{c}_j)
\end{eqnarray}
We can then write the marginalised posterior for $\kappa_s$ as
\begin{eqnarray} \label{eqn:clumppost2}
  P(\kappa_s|\vect{d}) = \frac{1}{Z_{\rm tot} N_c} \sum_{j=1}^{N_c} \int \mathcal{L}(\vect{d}|\vect{c}_j,\vect{\theta})\,P(\vect{\theta})\ d\vect{\theta}
\end{eqnarray}
Here $\kappa_s$ is implicit on the right-hand side because it determines the number of clumps.  Let us decompose this expression a little bit further.  We can think of the $\vect{\theta}$ integral as the Bayesian evidence for the macromodel, given a particular clump realisation, so let us put
\begin{eqnarray} \label{eqn:Zj}
  Z_j(\kappa_s) \equiv \int \mathcal{L}(\vect{d}|\vect{c}_j,\vect{\theta})\,P(\vect{\theta})\ d\vect{\theta}
\end{eqnarray}
We can then rewrite eqn.~(\ref{eqn:clumppost2}) as
\begin{eqnarray}
  P(\kappa_s|d) = \frac{1}{Z_{\rm tot} N_c} \sum_{j=1}^{N_c} Z_j(\kappa_s)
\end{eqnarray}
In other words, the marginalised posterior for $\kappa_s$ is the average macromodel evidence over many clump realisations (up to an overall normalisation factor).

In principle, we just need to use nested sampling to evaluate the $\vect{\theta}$ integral for each of many realisations, and then take the average.  There are two practical issues.  First, there is some statistical uncertainty in the average due to having a finite number of realisations.  We set $N_c=5000$ in order to sample the distribution well enough to achieve a statistical uncertainty of $\sim\!20\%$ in the evidence marginalised over clump realisations (verified with jackknife estimates).

Second, in our current implementation it can take several hours or more to run the $\vect{\theta}$ nested sampling for a given clump realisation, so it is impractical to do the full evidence calculation for all $9 \times 5000$ cases.  Instead, we explore the possibility of using the minimum $\chi^2$ value (which is much easier to determine) as a proxy for the evidence in each case.  The minimum $\chi^2$ provides a measure of the peak log likelihood, so it may be more or less indicative of the evidence depending on whether the width of the likelihood distribution is fairly regular or irregular.  For a subset of clump realisations we find both the best $\chi^2$ (by optimising across $\vect{\theta}$) and the full evidence (by integrating over $\vect{\theta}$), and we compare the two quantities in Figure \ref{fig:chievid}.

\begin{figure*}
\centering
\includegraphics[clip=true, trim=0cm 1cm 1cm 0.cm,width=15cm]{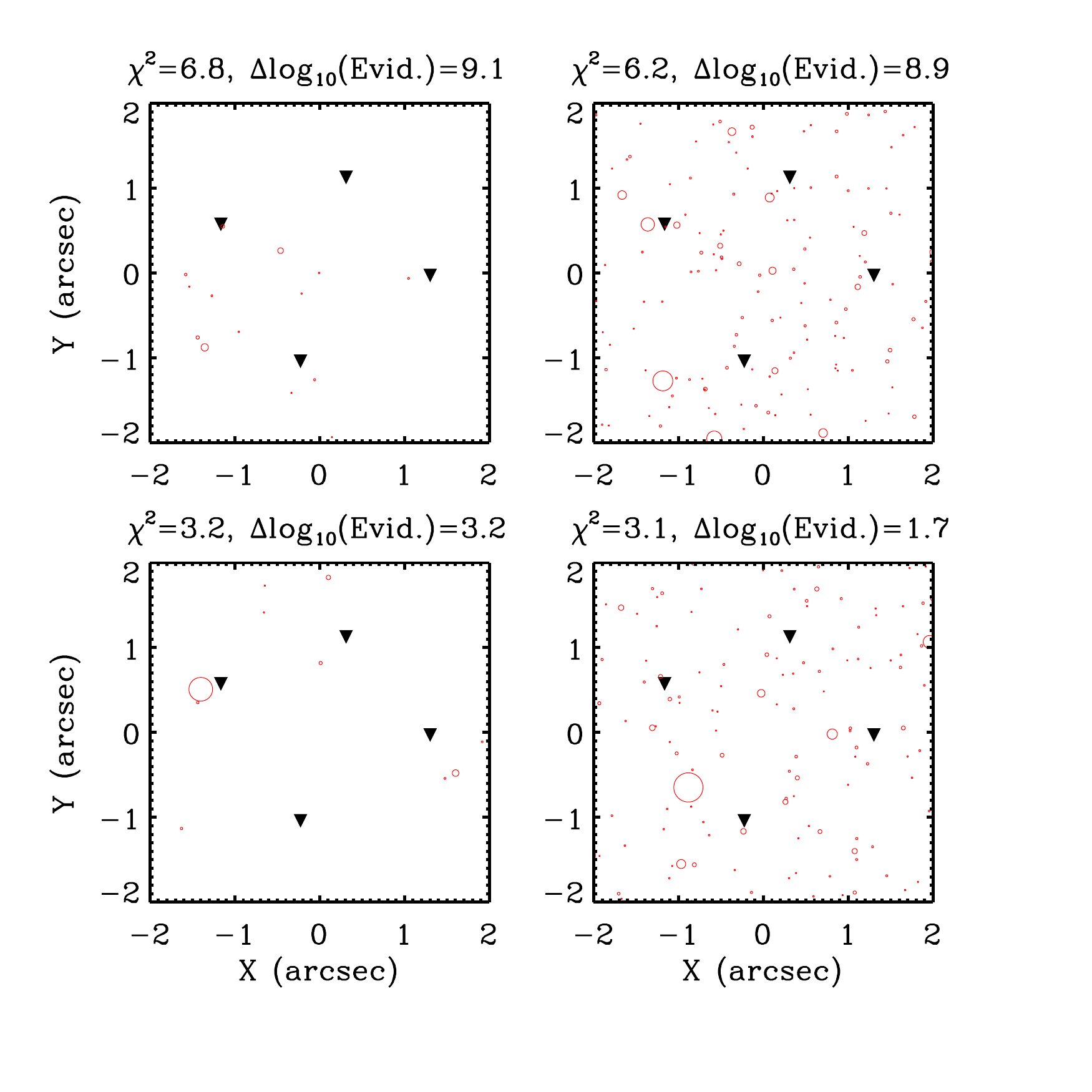}
\caption{
Spatial distributions of clumps for two realisations with $\kappa_s=0.001$ (left) and $\kappa_s=0.01$ (right).  Triangles mark the image positions, while circles indicate clumps with the circle size proportional to the clump Einstein radius.  Each realisation shown here provides a reasonable fit to the data and has at least one clump near image A (with spatial locations similar to our few-clump models).  Note, however, that the realisations in the top row have worse $\chi^2$ values but much higher evidence values than the realisations in the bottom row.  We conjecture that when there is a massive clump near the images (as in the bottom row) the smooth component is confined to a smaller region of parameter space, leading to a narrower posterior distribution and hence a lower evidence value.
}\label{fig:clumpdists}
\end{figure*}

Looking first at the case with $\kappa_s=0.001$, we see distinct patterns in the points.  At values $\chi^2\la6.5$, we find that $\chi^2$ is a very poor predictor of evidence: realisations with similar $\chi^2$ values can have evidence values that span many orders of magnitude.  This presumably occurs because certain realisations can produce good fits only for a highly tuned set of macro parameters, meaning the likelihood distributions are narrow in $\vect{\theta}$ and the evidences are small; whereas other realisations can have a much larger range of macro parameters that produce reasonable fits.  (See Fig.~\ref{fig:clumpdists} for more discussion.)  Above $\chi^2\sim6.5$, there is a tighter relation between $\chi^2$ and evidence.  The evidence decreases with $\chi^2$ up to $\chi^2\sim21$, at which point the evidence values become consistent with the evidence for the smooth, minimal model.  The patterns generally persist as $\kappa_s$ increases, although with more scatter.  We attribute the scatter to having more clumps and thus a wider range of substructure perturbations, some of which make the model more consistent with the data but many of which go in the opposite direction.

In Figure \ref{fig:chievid} we also plot the mean and scatter in the evidence values for bins of $\chi^2$.  We use these to create a ``lookup'' scheme to convert from $\chi^2$ to evidence for subsequent realisations.  Specifically, for each realisation we optimise across the macro parameters $\vect{\theta}$ to find the best-fit $\chi^2$.  We interpolate between bins to find the mean and scatter in the log evidence for that $\chi^2$ value.  We then draw from the appropriate log-normal distribution to assign an evidence value to this realisation.  With this process it becomes feasible to complete 5000 realisations for each value of $\kappa_s$.

During the course of this analysis, we initially found that some clump realisations led to unreasonably large best-fit values for the ellipticity of the neighbor galaxy G22 ($e_{\rm G22}\sim0.9$).  In order to prevent this, we adopted a mild Gaussian prior of $0.0\pm0.2$ on the two quasi-Cartesian ellipticity components.  While ad hoc, this prior is tight enough to prevent unrealistic values for the ellipticity and to stabilise the $\chi^2$ values, yet broad enough to allow a large range of ellipticity.

%=====================================================================
\subsection{Results}
\label{sec:pop-results}

\begin{figure}
\centering
\includegraphics[clip=true, trim=1.9cm 12.4cm 2.1cm 3.cm,width=8cm]{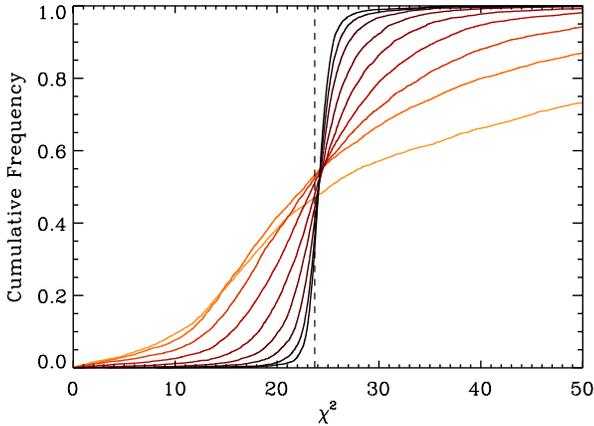}
\caption{
For a given $\kappa_s$, we generate many substructure realisations, find the best $\chi^2$ for each one, and plot the cumulative distribution of the resulting $\chi^2$ values.  Different colours indicate different $\kappa_s$ values ranging from 0.00022 (black) to 0.10 (light orange).  The vertical dashed line marks the optimised $\chi^2$ for our smooth, minimal model.  As $\kappa_s$ increases, the $\chi^2$ distribution broadens because some substructure realisations improve the fit while others worsen it.
}\label{fig:chidist}
\end{figure}

Figure \ref{fig:chidist} shows the cumulative distribution of $\chi^2$ values for different values of $\kappa_s$.  As $\kappa_s$ increase the distribution of $\chi^2$ values gets broader; in other words, with more substructure there is a higher chance that images will be perturbed and the model will move away from the smooth case.  The scatter goes in both directions: some clump realisations make the fit better, while others make it worse.  That is in the nature of stochastic substructure.

\begin{figure}
\centering
\includegraphics[clip=true, trim=3.0cm 12.8cm 2.1cm 3.cm,width=8cm]{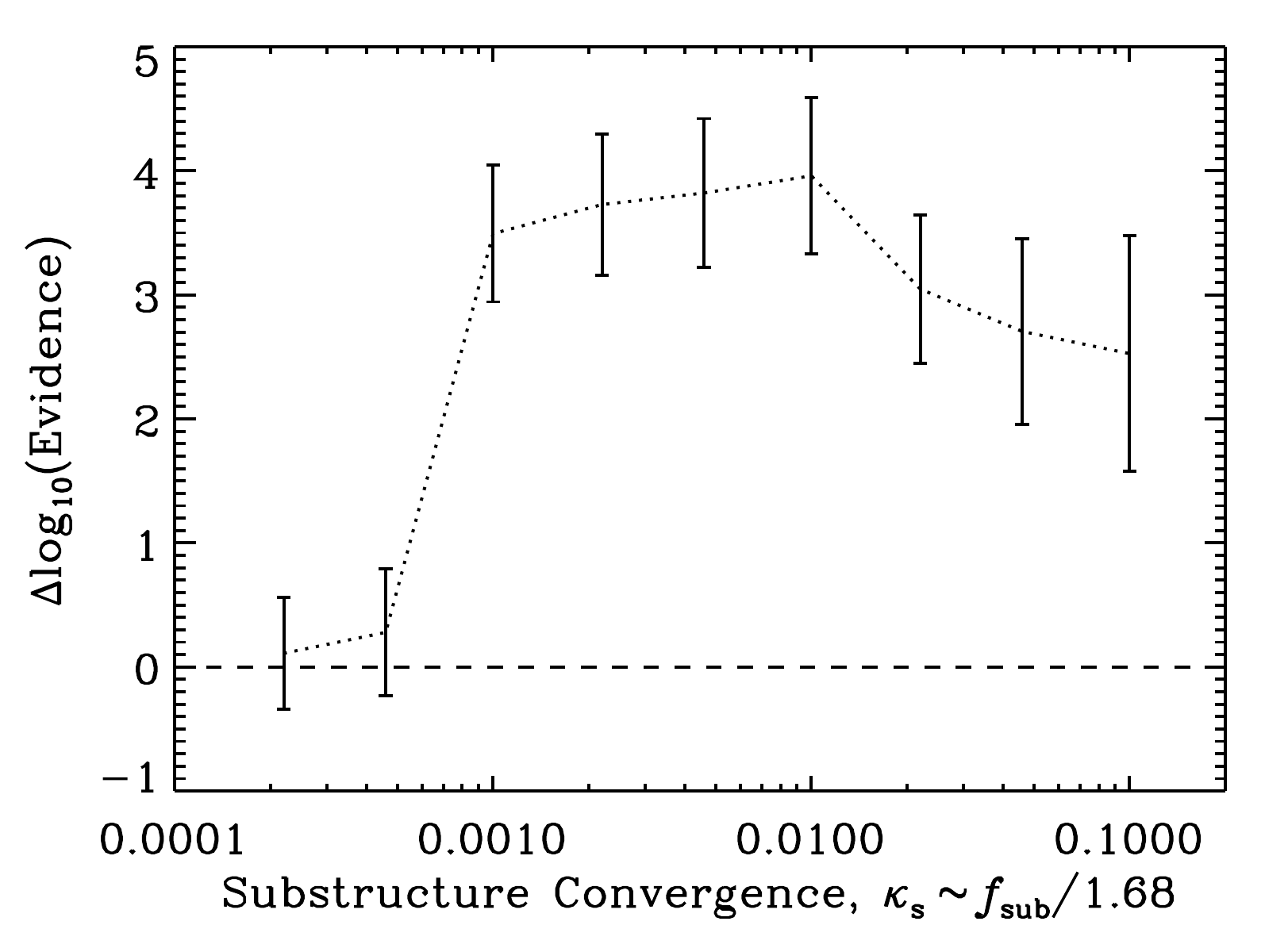}
\caption{
Differential log evidence (relative to the smooth, minimal model) as a function of the substructure convergence.  Models with $\kappa_s \le 0.00046$ exhibit evidence values similar to those without substructure.  Models with $\kappa_s\ge0.001$ are strongly favoured.  According to the Jeffreys' scale (Table \ref{tab:jeffreys}), these results provide decisive evidence for substructure in HE0435.
}\label{fig:fracevid}
\end{figure}

Figure \ref{fig:fracevid} shows the average evidence as a function of the substructure convergence.  We find that models with $\kappa_s\ge 0.001$ have evidence values that are some three orders of magnitude higher than models with little or no substructure.  According to the Jeffreys' scale, this is additional strong evidence for substructure in HE0435.  Moreover, we find that the $\Delta\log_{10}(\mbox{evidence})$ values for population models are similar to those for few-clump models (Table \ref{tab:results}), indicating that the data are consistent with, but do not objectively favour, millilensing by a full population of clumps.

We can translate the substructure convergence, $\kappa_s$, into a substructure mass fraction at the Einstein radius.  Our power law macromodels have $\kappa_0 \approx \beta/2$ at the Einstein radius \citep[see, e.g.,][]{kochanek02b}, so the local substructure mass fraction at the Einstein radius is $f_{\rm sub} = \kappa_s/\kappa_0 \approx 2\kappa_s/\beta$.  As noted in Section \ref{sec:few-results}, we find $\beta \approx 1.19$ with $\sim$14\% uncertainty.  Thus, given that we can decisively rule out values $\kappa_s \le 0.00046$, we conclude that $f_{\rm sub}>0.00077$ in HE0435 at high confidence.  At present we do not obtain an upper limit on $f_{\rm sub}$; the Bayesian evidence remains high for values as large as $f_{\rm sub}\approx0.20$.  We speculate that such high substructure mass fractions are permissible thanks to the flexibility and freedom available to our macro model.  

Our lower bound $f_{\rm sub}>0.00077$ is consistent with, but weaker than, other lensing-based measurements.  Using a sample of seven radio quads, \citet{dalal02a} found $0.006<f_{\rm sub}<0.07$ at 90\% confidence.  In the case of SDSS 0946+1006, \citet{vegetti10b} found $f_{\rm sub}=0.0215^{+0.0205}_{-0.0125}$ (68\% confidence), assuming $\alpha=-1.9\pm0.1$ for the slope of the substructure mass function and total mass thresholds $M_{\rm total}=10^{6.6}M_\odot$ and $10^{9.6} M_\odot$.  Both results point to values of $f_{\rm sub}$ that are higher than the values found in N-body simulations \citep[$f_{\rm sub}\sim0.002$--0.003; e.g.,][]{diemand07a,springel08a,xu10a}.  It is striking that HE0435 provides such strong evidence for substructure yet permits $f_{\rm sub}$ values that are low and fully consistent with CDM predictions.  This result might imply that HE0435 simply has less substructure than some other lenses.  Alternatively, it might indicate that something about HE0435 makes it less effective than some other lenses at constraining substructure.  It is important to remember that flux ratios are mainly sensitive to clumps near the images, so they directly probe a small fraction of the lens galaxy halo, whereas $f_{\rm sub}$ describes the global substructure population.  Regardless of whether lens galaxies have different amounts of substructure or simply different strengths of anomalies (due to the stochastic nature of substructure lensing), it is clear that future studies will need to examine ensembles of lenses (like the seven used by \citealt{dalal02a} but ideally even more) to produce strong, robust constraints on the global properties of dark matter substructure in galaxies.

Since our models permit, and other lensing results imply, $f_{\rm sub}$ values higher than CDM predictions, it is worth considering how such a discrepancy might arise and whether it presents a challenge to CDM.  On the theory side, one possible concern is that the number of surviving subhalos in N-body simulations might be underestimated, due to the lack of baryons in most simulations.  In general, baryons are expected to cause dark matter (sub)halos to contract and become more concentrated, and that may make subhalos less susceptible to tidal disruption \citep[e.g.,][]{dolag09a}.  On the other hand, however, baryons also make the parent halo more concentrated, and that might raise the rate of tidal disruption \citep[e.g.,][]{romanodiaz10a}.  Other possible concerns lie on the lensing side.  Currently, the number of lenses available for studying $f_{\rm sub}$ is small (of order 10) and thus sensitive to both statistical uncertainties and selection effects.  Lensing is generally biased toward more massive and concentrated galaxies, perhaps with preferential orientations along the line of sight \citep[e.g.,][]{rozo07a,mandelbaum09a}.  Furthermore, lens galaxies tend to lie in overdense environments \citep[e.g.,][]{momcheva06a}, and the environment may boost the abundance of substructure \citep{oguri05a}.  More work needs to be done to understand whether selection effects in lensing propagate into a significant bias in $f_{\rm sub}$.

Beyond our quantitative constraints on $f_{\rm sub}$, there are two aspects of our analysis worth highlighting.  First, we have examined both individual clump models and full substructure population models for a given system, and sought to connect them.  This is a first for quasar lenses.  For galaxy-galaxy strong lenses, \citet{vegetti10b} have detected a clump via gravitational imaging, and used that to infer $f_{\rm sub}$ by an analysis similar in concept to what we present in Appendix A.

Another key feature of our analysis is the method used for studying substructure populations.  Here, the only comparable study is that of \citet{dalal02a}.  Due to the complexity and computational demand of the study, \citeauthor{dalal02a} chose to linearly reoptimise the macromodel and then work with the best $\chi^2$ for each substructure realisation.  By contrast, we have fully marginalised the macromodel and worked with the Bayesian evidence.  We have found that the best $\chi^2$ value can be an unreliable tracer of the evidence, at least in the HE0435 system.  If such behavior occurs in the systems and models studied by \citeauthor{dalal02a}, it could affect the recovered $f_{\rm sub}$ values.  Additionally, \citeauthor{dalal02a} assumed a uniform mass for their substructure population.  Here, we have considered a more realistic population with a mass function like that seen in N-body simulations.  Due to the computational demand of our simulations, we have so far only examined one value of $\alpha$ and one range of clump masses.  Future work will explore the dependence of inferred $f_{\rm sub}$ values on these parameters.

%=====================================================================
\section{Implications}
\label{sec:implications}
%=====================================================================

While we have focused on using image positions and flux ratios to probe substructure in HE0435, our models have additional implications that we explore in this section.

%=====================================================================
\subsection{$K$-band flux ratios}
\label{sec:microK}

Our substructure models are able to account for the observed $R$-band flux ratios and, due to their similarity, the $L'$ flux ratios as well.  As noted in Section \ref{sec:constraints}, however, the $B/C$ flux ratio is a factor of 1.27 higher in $K$ than in the other passbands.  This anomaly is perplexing because the $K$-band emission (rest-frame 0.82 $\mu$m) presumably originates from the quasar accretion disk, so the $K$ and $R$ sources should both be small compared with the Einstein radii of dark matter clumps and should therefore experience similar lensing magnifications.

Since differential dust extinction is not likely in HE0435 \citep{wisotzki03a}, we hypothesise that the $K$-band anomaly is caused by stellar microlensing that happened to be stronger at the time of the infrared observations than during the earlier optical monitoring by \citet{kochanek06a}.  Under this hypothesis, the (nearly) contemporaneous $L'$ flux ratios were not anomalous because the $L'$ emission originates from a larger region that is less susceptible to microlensing.  We test whether this hypothesis is reasonable by simulating microlensing near image B using the ray-shooting code of \citet{wambsganss99a}.

We simulate microlensing in a box with side length $L=15\Rein$, where $\Rein$ is the Einstein radius for the mean star mass.  This box is chosen to be large compared with the source, but small enough that we can apply a uniform convergence and shear across the box.  We set the convergence and shear to the values predicted by our best-fit two-clump (AB) lens model: $\kappa_{\rm local} = 0.694$ and $\gamma_{\rm local} = 0.486$.  To divide $\kappa_{\rm local}$ into the contribution from stars ($\kappa_\star$) and a contribution that is smooth on the scale of the box (i.e., from dark matter), we use results from star+dark matter lens models for HE0435 by \citet{kochanek06a}.  Those models yield $\kappa_\star=0.05\log_{10}(r_c h/{\rm kpc})$, where $r_c$ is the NFW scale radius in the models.  For the range of values, $2.5'' < r_c < 20''$, considered by \citet{kochanek06a}, the $\kappa_\star$ values lie between 0.05 and 0.10.  We try both extremes.

For each value of $\kappa_\star$ we generate 100 random realisations of the stellar distribution, drawn from a mass function $dN/dm \propto m^{-1.3}$ over the range 0.01--1.5 $M_\odot$.  Such a mass function agrees with measurements from the Galactic bulge \citep{gould00a} and has been used in previous microlensing studies \citep{morganc10a, poindexter10a}.  With this mass function and the source/lens redshifts appropriate for HE0435, the Einstein radius for the mean stellar mass is $\log_{10}(\Rein/'')=-6.1$.

For each realisation we use ray shooting to construct a magnification map with resolution $L/1024$.  We then convolve the map with a uniform circular source whose size corresponds to the expected size of the $K$-band emission region.  To estimate the source size, we start with the empirical results for $I$-band sources from \citet{morganc10a}, and then scale from $I$ (rest-frame 0.26 $\mu$m) to $K$ using the familiar relation $R \propto \lambda^{4/3}$ from \citet{shakurya73a}.  This yields a $K$-band source size of $\log_{10}(R_{\rm src}/'')=-6.1^{+0.5}_{-0.7}$, or $R_{\rm src}/\Rein = 0.2$--3.1 (68\% CL).  Already we see that the $K$-band source has a size that should make it sensitive to microlensing.

\begin{figure}
\centering
\includegraphics[clip=true, trim=2.2cm 12.4cm 2.0cm 3.2cm,width=8cm]{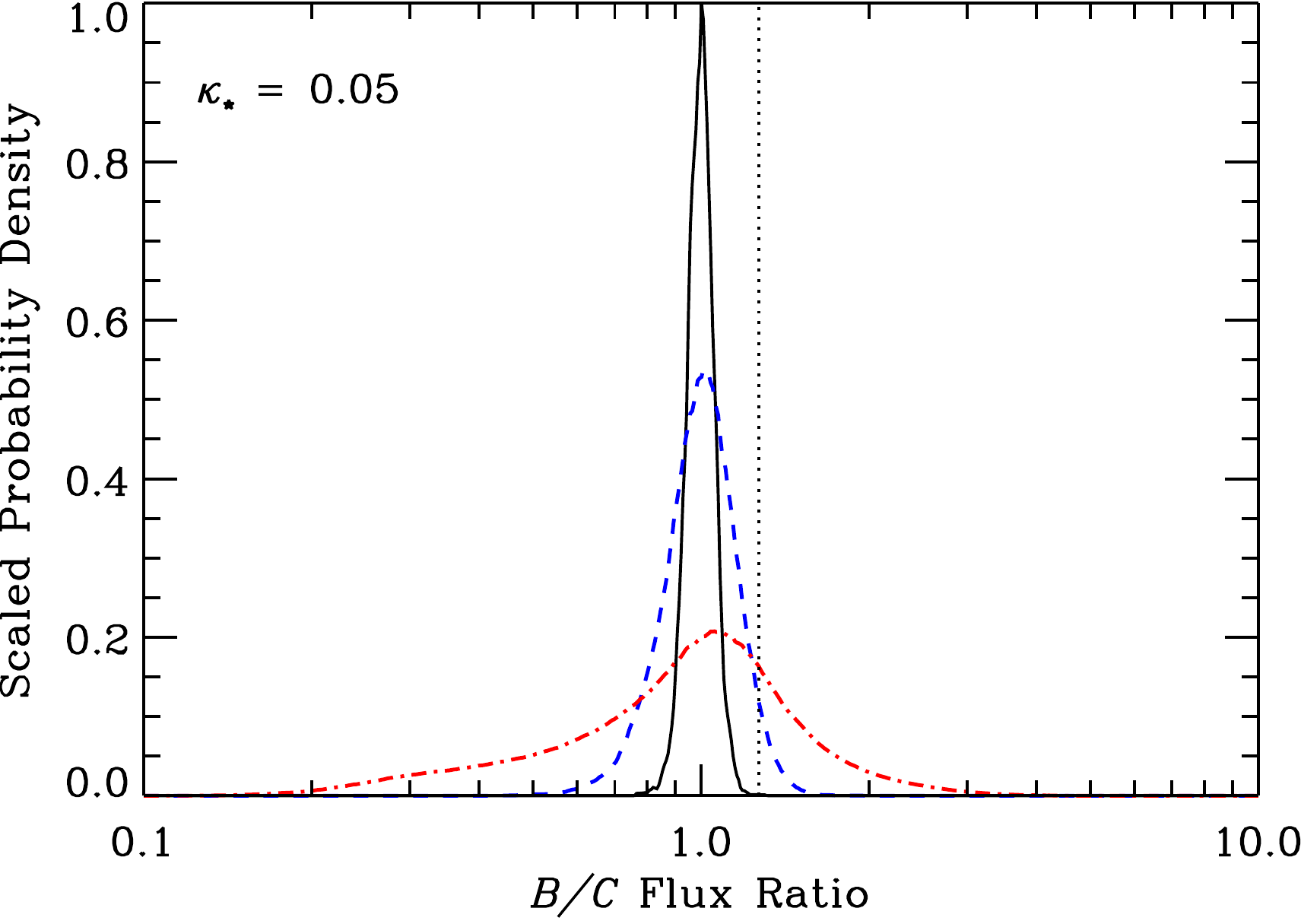}
\includegraphics[clip=true, trim=2.2cm 12.4cm 2.0cm 3.2cm,width=8cm]{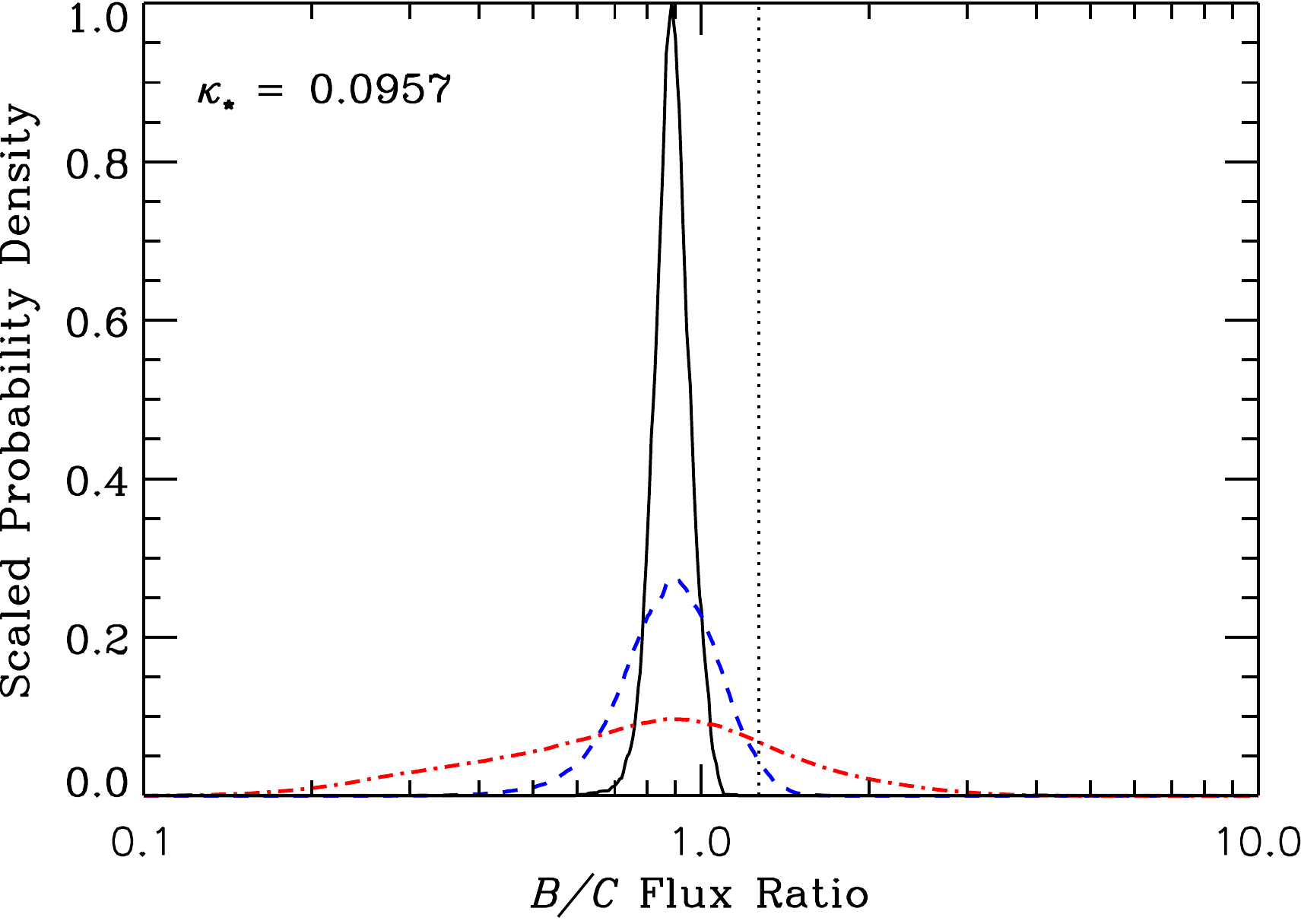}
\caption{
Probability distributions for the $B/C$ flux ratio under the influence of microlensing.  The three curves correspond to different source sizes: median size (dashed/blue), 68\% CL high value (solid/black), and 68\% CL low (dot-dashed/red), using the size distribution from \citet{morganc10a}.  The vertical dotted line indicates the value needed to reproduce our $K$-band observations.  We find that small source sizes $\la 0.48$ or 0.73 (in units of the Einstein radius of the mean star mass) can reproduce the observations well for $\kappa_*=0.05$ or 0.10, respectively.
} \label{fig:micro}
\end{figure}

Figure \ref{fig:micro} shows results from the microlensing simulations.  In general, it is not hard for microlensing to increase the magnification of image B by a factor of 1.27.  To go into more detail, we consider what range of source sizes can yield a microlensing boost $>1.27$ more than 16\% of the time (corresponding to a two-sided 68\% CL).  This requires a source size of $R_{\rm src}/\Rein \la 0.48$ (0.73) for $\kappa_\star=0.05$ (0.10).  Such sizes are well within the estimated range for the $K$-band source ($R_{\rm src}/\Rein \sim 0.2$--3.1), so it seems quite plausible that microlensing can create the additional magnification necessary to explain the observed $K$-band flux ratio.

At the same time, we must consider whether microlensing would affect the $L'$ flux enough to disturb the agreement between the model and data.  In a ``worst case'' scenario we might imagine that all the $L'$ emission originates from the accretion disk.  In this case the $L'$-band source size would need to be $R_{\rm src}/\Rein \ga 1.0$--1.3 to keep the predicted flux ratio consistent with the observed value.  With the \citeauthor{shakurya73a} scaling this would imply an associated $K$-band source size of $R_{\rm src}/\Rein \ga 0.54$--0.63, which is compatible with the range of values needed to reproduce the observed $K$-band flux ratio.  Since it is likely that some of the $L'$ emission originates from the larger torus, which would be even less sensitive to microlensing, we infer that the observed $L'$-band flux ratio is not inconsistent with significant microlensing in the $K$-band.

Given uncertainties in the source sizes, stellar density, stellar mass function, and measured flux ratios, we conclude that microlensing provides a plausible explanation for the observed $B/C$ flux ratio in the $K$-band.  Confirmation of this hypothesis will be possible with future $K$-band observations to quantify variability in the flux ratios.

%=====================================================================
\subsection{Time delays}
\label{sec:delays}

After our modeling was complete, \citet{courbin10a} presented new data for HE0435 including a measurement of the velocity dispersion of the lens galaxy and new estimates of the time delays between the images derived from six years of photometric monitoring.  The new time delays have values similar to those obtained by \citet{kochanek06a} but errorbars that are a factor of $\sim$2.6 larger.  The primary origin of the increased uncertainties lies in the analysis methods used by the two teams.  In particular, \citet{courbin10a} find a larger uncertainty in the arrival time of image A.

There is no obvious way to post-process our modeling results to apply the new time delay constraints in a rigorous fashion.  Nevertheless, we can compare the distribution of time delays predicted by our models with the new measurements.  This test is statistically meaningful because we did not place any constraints on time delays when fitting the models.

\begin{figure*}
\centering
\includegraphics[clip=true, trim=2.1cm 12.05cm 2.5cm 2.75cm,width=8cm]{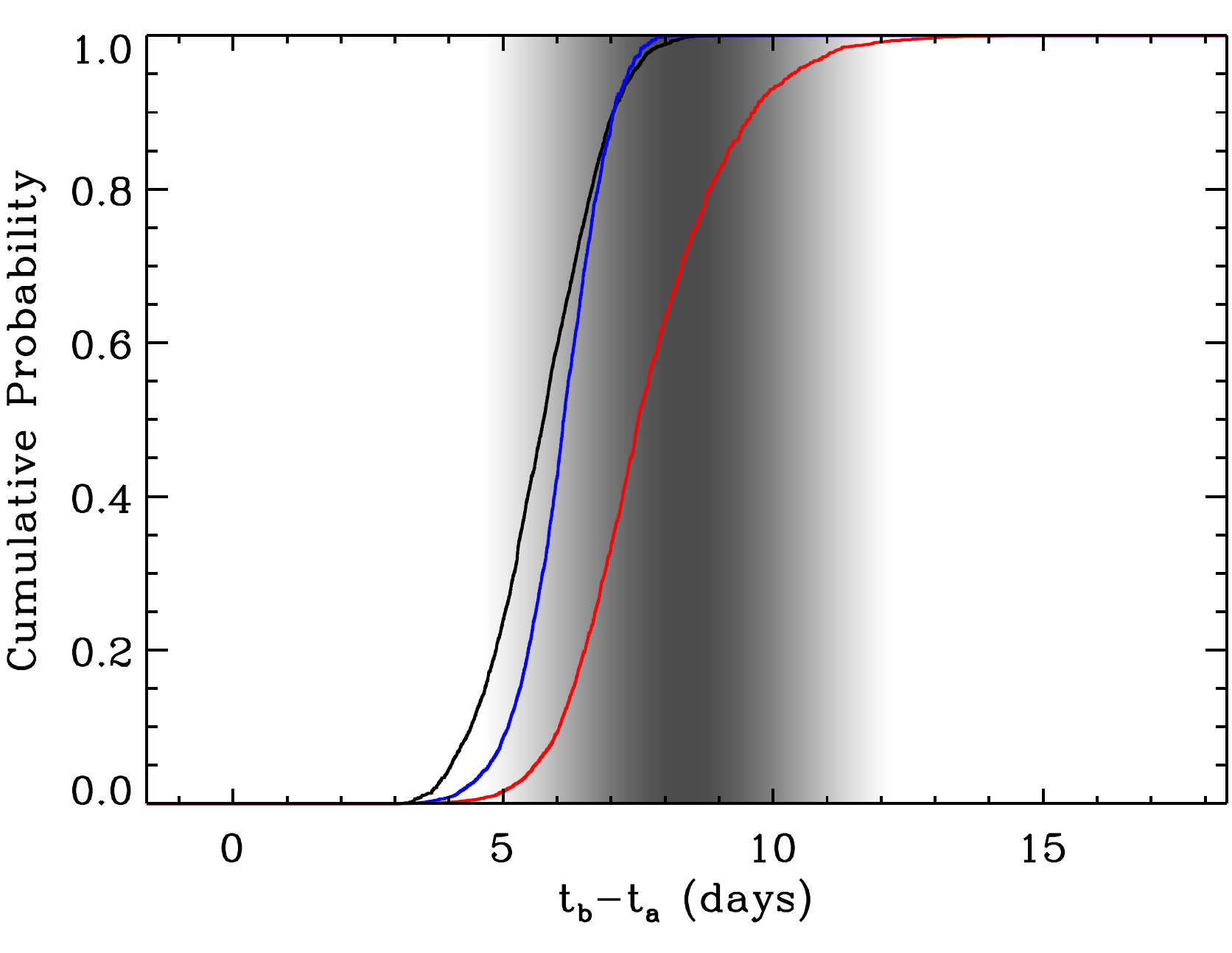}
\includegraphics[clip=true, trim=2.1cm 12.05cm 2.5cm 2.75cm,width=8cm]{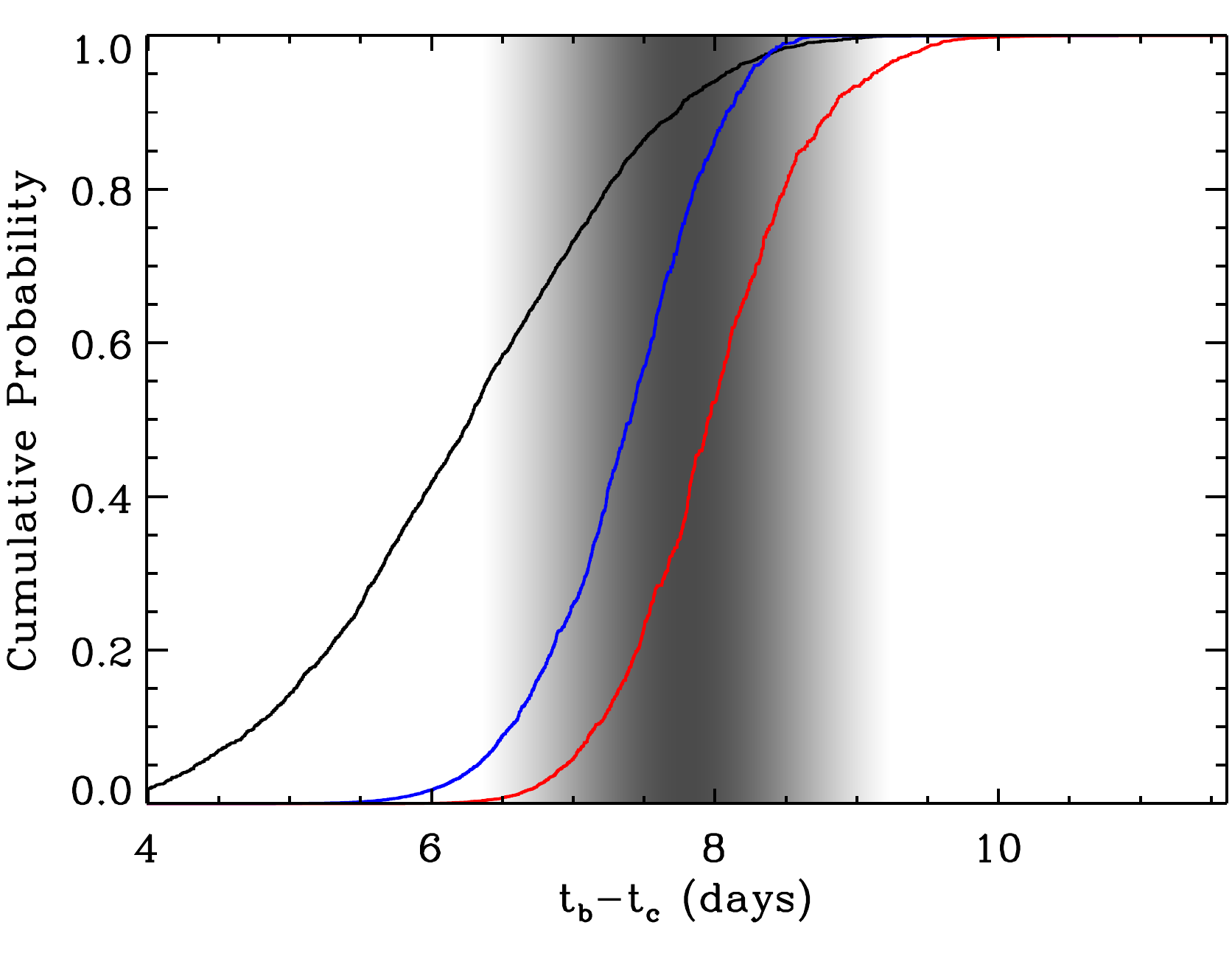}
\includegraphics[clip=true, trim=2.1cm 12.05cm 2.5cm 2.75cm,width=8cm]{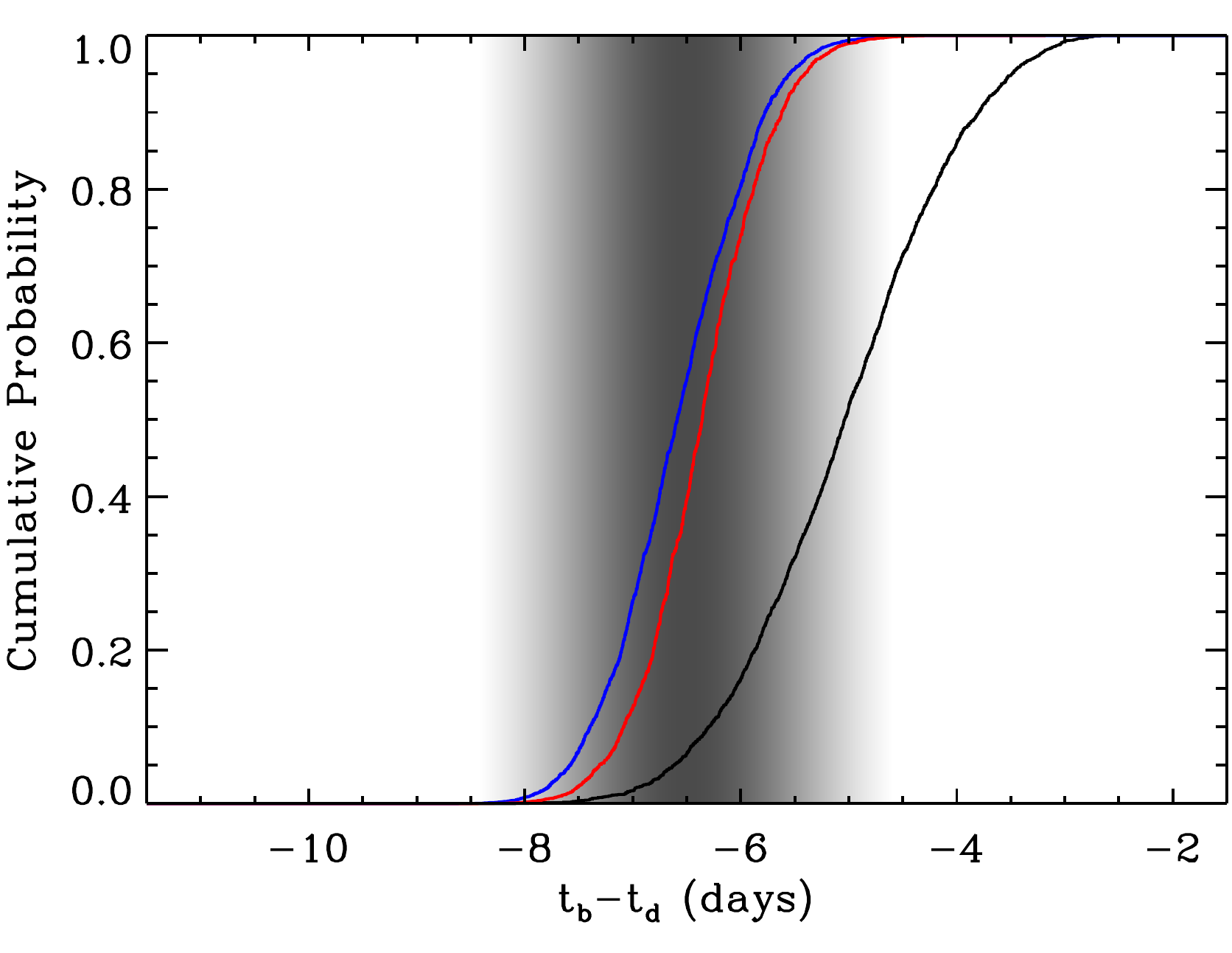}
\caption{
Cumulative probability distributions of the time delays predicted by three models: smooth (black), smooth + clump near A (blue), and smooth + clumps near A and B (red).  The shaded grayscale indicates the newly-measured time delays from \citet{courbin10a}.  We find that models with substructure generally predict time delays that agree with the data, even though the observed values were not used as constraints, whereas models without substructure do less well.  Future studies will benefit from adding time delay constraints.
}\label{fig:tdels}
\end{figure*}

Figure \ref{fig:tdels} shows the comparison.  We find that models without substructure are somewhat discrepant from models that include a few clumps.  Furthermore, the predictions from our few-clump models are in good agreement with the new time delays measurements.  It is clear, though, that adding time delay constraints would further improve the models.  As a check, we have also examined predicted time delays for our population models (not shown).  These, too, are in good agreement with the new data, and may benefit even more from new constraints because \emph{populations} of substructure tend to broaden time delay distributions \citep[see][]{keeton09a}.  The improvement would be particularly relevant for models with high $f_{\rm sub}$ values, since time delays scatter with $\sigma_\tau\propto\sqrt{f_{\rm sub}}$.  We plan to include the new time delays, as well as velocity dispersion constraints, in future work.

%=====================================================================
\section{Conclusions}
\label{sec:conclusions}
%=====================================================================

We have conducted a fully Bayesian analysis of the lens HE 0435$-$1223.  Combining astrometry from HST with flux ratios from ground-based monitoring, we probe the mass distribution down to milli-arcsecond scales, and assess various models for substructure using the Bayesian evidence.  We also examine multi-wavelength properties of the lens.  We summarise our conclusions as follows:

\begin{itemize}

\item The observed flux ratio of images A and C cannot be reproduced by macroscopic, smooth lens models.  This failure cannot be due to microlensing or intrinsic variability of the background quasar, since monitoring has quantified such variations.  Instead, we find that adding a single clump near image A whose mass within the Einstein radius is $\log_{10}(\Mein^A)=7.65^{+0.87}_{-0.84}$ (in units of $h_{70}^{-1}\,M_\odot$) can account for the data.

\item We consider other sources of substructure by including additional clumps near other lensed images.  Using the Bayesian evidence to compare the various possibilities, we find that a model with clumps near images A and B is most favoured.  This model has an evidence that is 0.63 dex greater than our single clump model, and implies a mass for clump B of $\log_{10}(\Mein^B)=6.55^{+1.01}_{-1.51}$.  However, the modest increase in the evidence, coupled with evidence uncertainties, makes the case for clump B less decisive than the case for clump A.

\item We also examine a full ensemble of subhalos, using a mass function consistent with CDM predictions ($dN/dM \propto M^{-1.9}$ over the range $10^7$--$10^{10}\,M_\odot$) and varying the abundance of substructure.  By examining the Bayesian evidence, we infer the mass fraction of substructure to be $f_{\rm sub}>0.00077$ near the Einstein radius.  Our measurement of $f_{\rm sub}$, unlike other lensing-based measurements, is fully consistent with that predicted by CDM simulations ($f_{\rm sub} \sim 0.002$--0.003).

\item As part of our substructure analysis, we find that \emph{optimising} the macromodel for each realisation of substructure cannot necessarily substitute for \emph{marginalising} the macromodel.  In the Bayesian framework, full marginalisation is important.

\item Near-infrared flux ratio measurements in the $K$ and $L'$ passbands generally agree with those from optical monitoring.  The lone exception is the $K$-band value of the flux ratio $B/C$.  We show that stellar microlensing provides a plausible explanation for the $K$-band flux of image B, if the $K$-band source has size $\log_{10}(R_{\rm src}/'') \la -6.24$.  Estimates based on accretion disk measurements by \citet{morganc10a} 
suggest that such a size is reasonable.  Future $K$-band observations can test the microlensing hypothesis.

\end{itemize}

%=====================================================================
\section*{Acknowledgments}

We thank Phil Marshall for a very helpful and insightful referee report, and Dominique Sluse for interesting discussions about HE0435.
Work presented here received support from the US National Science Foundation through grant AST-0747311.

%=====================================================================
\appendix
\section{Connecting individual clumps to the population}

In Section \ref{sec:pop} we specify the form of the mass function and spatial distribution from which our substructure population is drawn.  The free parameter in our analysis is the amount of substructure, characterised by $\kappa_s$.  Here we present a toy model to extrapolate from our few-clump models (Section \ref{sec:few}) and estimate $\kappa_s$.

For clumps near an image, let $A(m)$ be the area of the ``region of influence'' for a clump with scaled mass $m$ to produce a perturbation.  Focusing on magnification perturbations, we have $A(m)\propto m$ since magnification perturbations are driven by shear perturbations of the form $\delta\gamma \propto m/d^2$ where $d$ is the distance of the clump from the image (see Section \ref{sec:few-results}).  We can set the proportionality constant using our few-clump models: if there is a clump with scaled mass $m_i$ at distance $d_i$ from image $i$, then we put
\begin{eqnarray}
  A(m)=\frac{A_i}{m_i}\,m
\end{eqnarray}
where for simplicity we let $A_i$ be the geometric area $\pi d_i^2$.

Under these assumptions, the mean number of clumps that lie within the region of influence for image $i$ is
\begin{eqnarray} \label{eqn:lambdai}
  \lambda_i = \int A(m)\,\frac{dn}{dm}\ dm = \frac{A_i}{m_i}\,\kappa_{s,i}
\end{eqnarray}
where $\kappa_{s,i}$ is the substructure convergence near image $i$ (qv.\ eqn.~\ref{eqn:kappas}).  In our main analysis we assume a uniform substructure convergence, so $\kappa_{s,i}=\kappa_s$ for all images, but this framework could be made more general.  The probability that there are $N_i$ clumps affecting the image is a Poisson distribution of the form 
\begin{eqnarray}
  P(N_i|\lambda_i)=\frac{\lambda_i^{N_i}e^{-\lambda_i}}{N_i!}\ .
\end{eqnarray}
Inspired by our few-clump models, we consider the probability that there is one clump affecting image A:
\begin{eqnarray} \label{eqn:Pclump1}
  P(1|\lambda_A) = \lambda_A e^{-\lambda_A} .
\end{eqnarray}
Since $\lambda_A$ depends on $\kappa_s$, we can reinterpret this as a probability distribution for $\kappa_s$.  Specifically, in a Bayesian framework with flat priors, the posterior distribution for $\kappa_s$ has the form
\begin{eqnarray} \label{eqn:Pclump2}
  P_{\rm clump}(\kappa_s | m_A, A_A) \propto \frac{A_A}{m_A}\,\kappa_s \exp\left( -\frac{A_A}{m_A}\kappa_s \right)
\end{eqnarray}
where $m_A$ is the mass of the clump near image A, and $A_A=\pi d_A^2$ is the area defined by the distance $d_A$ of the clump from the image.  This distribution has a peak at $\kappa_s=m_A/A_A$, a mean of $\langle\kappa_s\rangle=2m_A/A_A$, and a standard deviation of $\sigma_{\kappa_s}=\sqrt{2} m_A/A_A$.  In practice we account for uncertainties in $m_A$ and $A_A$ by integrating over the allowed range,
\begin{eqnarray}
  P_{\rm clump}(\kappa_s) = \sum_{(m_A,A_A)} P_{\rm clump}(\kappa_s | m_A, A_A)\ 
    P(m_A, A_A)
\end{eqnarray}
where $P(m_A, A_A)$ is the posterior probability for the parameter pair $(m_A, A_A)$, and the sum runs over allowed pairs.  Using this ``final'' $P_{\rm clump}(\kappa_s)$ curve we find the median value and 95\% confidence range $\kappa_s = 0.025^{+0.074}_{-0.022}$.

%=====================================================================

\bibliographystyle{mn2e}

\end{document}